\documentclass[aps,11pt,onecolumn,preprintnumbers,amsmath,amssymb]{revtex4}

\usepackage[english]{babel}
\usepackage{graphicx}
\usepackage{color}
\usepackage{float}
\begin{document}

\newcommand{\unit}[1]{\:\mathrm{#1}}            
\newcommand{\To}{\mathrm{T_0}}
\newcommand{\Tp}{\mathrm{T_+}}
\newcommand{\Tm}{\mathrm{T_-}}
\newcommand{\EST}{E_{\mathrm{ST}}}
\newcommand{\Rp}{\mathrm{R_{+}}}
\newcommand{\Rm}{\mathrm{R_{-}}}
\newcommand{\Rpp}{\mathrm{R_{++}}}
\newcommand{\Rmm}{\mathrm{R_{--}}}
\newcommand{\ddensity}[2]{\rho_{#1\,#2,#1\,#2}} 
\newcommand{\ket}[1]{\left| #1 \right>} 
\newcommand{\bra}[1]{\left< #1 \right|} 

\bibliographystyle{naturemag}

\title{Entanglement of single-photons and chiral phonons in atomically thin WSe$_2$}

\author{Xiaotong Chen$^{1,\dagger}$}
\author{Xin Lu$^{1,\dagger}$}
\author{Sudipta Dubey$^{1,\dagger}$}
\author{Qiang Yao$^1$}
\author{Sheng Liu$^2$}
\author{Xingzhi Wang$^2$}
\author{Qihua Xiong$^{2,3}$}
\author{Lifa Zhang$^4$}
\author{Ajit Srivastava$^{1,*}$}
\affiliation{$^1$Department of Physics, Emory University, Atlanta 30322, Georgia, USA}
\affiliation{$^2$Division of Physics and Applied Physics, School of Physical and Mathematical Sciences, Nanyang Technological University, Singapore 637371, Singapore}
\affiliation{$^3$NOVITAS, Nanoelectronics Centre of Excellence, School of Electrical and Electronic Engineering, Nanyang Technological University, Singapore, 639798, Singapore}
\affiliation{$^4$Department of Physics, Nanjing Normal University, Nanjing, Jiangsu 210023, China.}

\maketitle
*Correspondence to: ajit.srivastava@emory.edu

\textbf{Quantum entanglement is a fundamental phenomenon which, on the one hand, reveals deep connections between quantum mechanics, gravity and the space-time~\cite{MaldacenaFP2013, CaoPRD2017}; on the other hand, has practical applications as a key resource in quantum information processing~\cite{NielsenBook2011}. While it is routinely achieved in photon-atom ensembles~\cite{MoehringNature2007}, entanglement involving the solid-state~\cite{GaoNature2012,DeGreveNature2012,DoldeNPhys2013} or macroscopic objects~\cite{LeeScience2011} remains challenging albeit promising for both fundamental physics and technological applications. Here, we report entanglement between collective, chiral vibrations in two-dimensional (2D) WSe$_2$ host --- chiral phonons (CPs) --- and single-photons emitted from quantum dots~\cite{SrivastavaNNano2015,KoperskiNNano2015,ChakrabortyNNano2015,HeNNano2015,TonndorfOptica2015} (QDs) present in it. CPs which carry angular momentum were recently observed in WSe$_2$ and are a distinguishing feature of the underlying honeycomb lattice~\cite{ZhuScience2018,ZhangPRL2015}. The entanglement results from a ``which-way" scattering process, involving an optical excitation in a QD and doubly-degenerate CPs, which takes place via two indistinguishable paths. Our unveiling of entanglement involving a macroscopic, collective excitation together with strong interaction between CPs and QDs in 2D materials opens up ways for phonon-driven entanglement of QDs and engineering chiral or non-reciprocal interactions at the single-photon level.}

Two dimensional (2D) materials with a honeycomb lattice, such as graphene and WSe$_2$, have degenerate, low energy electronic states which exhibit handedness. In the momentum space, this handedness is identified with the valley pseudo-spin,  labeled by $\pm K$-points of the Brillouin zone which are related to each other by time-reversal symmetry~\cite{XuNPhys2014, XiaoPRL2012}. This feature, arising from the presence of two sublattices, is at the heart of ``valley physics" in 2D materials, and has been exploited to create chiral optical excitations using the helicity of incident light~\cite{MakNNano2012,ZengNNano2012,CaoNComm2012}. Similarly, it has been predicted~\cite{ZhangPRL2015}, and recently observed~\cite{ZhuScience2018}, that the lattice vibrational modes or phonons in these materials can also have handedness or chirality. Chiral phonons (CPs) in a honeycomb lattice exist both at the center and the boundary of the Brillouin zone, carrying a (pseudo-) angular momentum of $l$ = $\pm 1$ along the out-of-plane direction (Fig.~1a). CPs are particularly intriguing as they represent chiral, collective motion of a macroscopic number of atoms as opposed to the chirality of a single charge carrier or an optical excitation. The possibility of chirality dependent coupling between a single optical excitation and CPs raises the question whether quantum control of the former, which is routinely achieved, can lead to a similar quantum control of collective excitations. Indeed, quantum state preparation, especially entanglement, involving macroscopic objects is not only a major goal of quantum information technology but also of fundamental interest and is being actively pursued in a variety of physical systems~\cite{GaoNature2012,DiCarloNature2010,SteffenScience2006}. 

Here, we show experimental evidence for entanglement between CPs of monolayer WSe$_2$ and the corresponding phonon-scattered, single-photons emitted from an embedded QD. The CP modes involve a collective excitation of $\sim$ 10$^9$ atoms of the monolayer as compared to the single optical excitation of the QD. The observed entanglement arises due to (pseudo-) angular momentum (AM) selection rules~\cite{ZhangPRL2015} of the phonon-scattering process which correlates the polarization of the single photon with the AM of doubly degenerate chiral phonons (CPs) involved. In particular, an exciton in the QD undergoes phonon emission via two indistinguishable paths involving doubly-degenerate CPs with opposite AM. As a result, the state of the emitted photon-phonon system becomes maximally entangled. To demonstrate this entanglement, we analyze the polarization of the photon subsystem, discarding any information of the phonon AM. This results in randomization of the photon polarization which is determined to be in a completely mixed state with a fidelity of 99 $\pm$ 1.5 $\%$. We emphasize that randomization of polarization is not expected in a coherent scattering process as there is no way for the polarization information to be lost and arises here as a consequence of intrinsic lack of knowledge in a ``which-path" scattering. Indeed, the entanglement is destroyed and the polarization recovered once the ``which-path" indistinguishability is removed by an out-of-plane magnetic field which breaks the time-reversal symmetry. This further rules out possible stochastic nature of the emitted phonon polarization as the reason for randomization of photon polarization because such a mechanism should not be affected by magnetic field.

Fig.~1a shows the phonon spectra of monolayer WSe$_2$ calculated using Quantum Espresso code (see Methods). The optical phonon modes (LO/TO) at $\Gamma$-point with energy of $\sim$ 20 meV are orthogonal and doubly-degenerate. They transform as $E''$ and can be superposed to get CPs with AM of $l$ = $\pm$ 1~\cite{ZhangPRL2015}. The $l$ = +1 (-1) mode corresponds to a lattice vibration in which the $W$ atoms remains stationary while the $Se$ atoms rotate around them in a counterclockwise (clockwise) fashion. There are also finite momentum CPs at the $\pm K$-points carrying similar AM and were recently observed by ultrafast spectroscopy~\cite{ZhuScience2018}. Fig.~1b shows Raman spectra measured on WSe$_2$ samples of different thickness down to monolayer. Consistent with previous studies~\cite{LuoPRB2013, Kim2DMaterials2017}, we observe a Raman peak corresponding to CP, $E''(\Gamma)$, at 176 cm$^{-1}$ or 21.8 meV, which is in very good agreement with the calculations. The Raman peak is not observed in the monolayer sample due to symmetry considerations which forbid it (see Supplementary Information). 

Fig.~1c and 1d show the basic idea behind entanglement of a CP with an optical excitation. Consider an optical excitation with in-plane linear polarization ($\pi^x$/$\pi^y$); for example, arising from an anisotropic potential which defines a preferred direction of the in-plane polarization; coupled to CPs through electron-phonon coupling. The optical excitation can be considered to be a superposition of left and right circularly polarized ($\sigma^+$/$\sigma^-$) states which carry an angular momentum of $\pm$1 along the out-of-plane direction. Due to conservation of AM in a three-fold symmetric crystal~\cite{ZhangPRL2015}, the $l$ = +1 (-1) phonon couples only with $\sigma^-$ ($\sigma^+$) photon as shown in Fig.~1d (see Supplementary Information). In other words, due to the degeneracy of phonons, the photon states involved in the superposition can scatter by two indistinguishable paths with AM conservation correlating the states of their polarization. As a result, after the scattering process, the state of the phonon-photon is -
\begin{eqnarray}
|\Psi_{\mathrm{tot}}\rangle = \frac{1}{\sqrt{2}}\left[ |\sigma^+ \rangle \otimes | l_{\mathrm{phonon}} = +1 \rangle \pm i |\sigma^- \rangle \otimes |l_{\mathrm{phonon}} = -1 \rangle \right],
\end{eqnarray}
where $\pm$ follow from $\pi^x$/$\pi^y$ polarization of the incoming photon and $\sigma^+$ /$\sigma^-$ label the helicity of the out-going photon. This state is nothing other than a maximally entangled state of phonon-photon system.

To realize the entanglement outlined above, we make use of the fact that a localized neutral exciton ($X^0$) in a WSe$_2$ QD~\cite{BrannyNComm2017,PalaciosBerraqueroNComm2016,PalaciosBerraqueroNComm2017} is split into two orthogonal linearly polarized states due to anisotropic electron-hole (e-h) exchange interaction~\cite{SrivastavaNNano2015,KoperskiNNano2015,ChakrabortyNNano2015,HeNNano2015}. Indeed, this fine-structure splitting is a hallmark of localized $X^0$ in many QD systems including self-assembled GaAs QDs~\cite{GammonPRL1996}. The use of QDs should also enhance the electron-phonon coupling due to quantum confinement effects and lower the symmetry to allow CP-scattering even in monolayer crystals (see Supplementary Information). 

We perform polarization-resolved photoluminescence (PL) spectroscopy on a monolayer WSe$_2$ sample in a field effect transistor (FET) device at low incident powers (see Methods). As shown in Fig.~2a, at positive gate voltage (V$_\mathrm{g}$ $\sim$ +20 V) when the sample is electron-doped, we observe the attractive Fermi-polaron peak ($X_{\mathrm{ap}}$)~\cite{SidlerNPhys2016} while for -90 V $<$ V$_\mathrm{g}$ $<$ 20 V, a faint free-exciton peak ($X^0$) is visible. Around V$_\mathrm{g}$ = 20 V, the so-called defect peak ($D$) appears and dominates the PL signal. Remarkably, this coincidences very well with the appearance of sharp QD peaks at lower energies than $D$, showing that the QDs originate from the WSe$_2$ monolayer. 

Fig.~2b shows PL spectra focusing on a few QD-like peaks. At V$_\mathrm{g}$ $\sim$ -23 V, when the sample is depleted of electrons, two set of doublets labeled D3a and D3b appear simultaneously. Moreover, the two doublets spectrally wander in identical manner, as highlighted by solid symbols. Consequently, we identify D3a and D3b to be originating from the same QD, D3. Furthermore, D2a/D2b and D4a/D4b pairs of doublets display similar behavior in their turn-on voltage and spectral wandering and are assigned to QDs D2 and D4, respectively. Fig.~2c shows cross-sectional PL spectra at a fixed V$_\mathrm{g}$. We first notice that the energy splitting of the $a$-doublets is identical to that of the corresponding $b$-doublets. Moreover, the energy spacing between the $a$- and $b$-peaks for all the three QDs is precisely 21.8 meV. As this energy spacing is exactly equal to the measured energy of the $E''$($\Gamma$) phonon in Raman spectra (Fig.~1b), we conclude that the lower energy $b$-peaks are the chiral phonon-replica of the parent $a$-peaks. Moreover, the $a$-peaks being doublets is consistent with localized $X^0$ and the corresponding fine-structure splitting of $\sim$ 600 $\mu$eV is in excellent agreement with previous studies~\cite{SrivastavaNNano2015,KoperskiNNano2015,ChakrabortyNNano2015,HeNNano2015, TonndorfOptica2015}. We observe this behavior in several other QDs (see Supplementary Information).

To further confirm that the $b$-peaks are indeed phonon replicas and not other excitonic complexes such as biexciton~\cite{HeNComm2016} or evenly-charged excitons, we first perform excitation power-dependence of all the peaks as shown in Fig.~2d. All the peaks exhibit a sub-linear, power-law behavior with emission intensity ($I$) scaling with the incident power ($P$) as $P^\alpha$ with $\alpha$ $\sim$ 0.8, thus ruling out biexciton as a possible origin of the $b$-peaks. Fig.~2e shows that both $a$- and $b$-doublets display Zeeman splitting in an out-of-plane magnetic field ($B$) with almost identical g-factor as is expected for the parent and phonon replica peaks. The measured value of g-factor ($\sim$ 9) is consistent with previous studies of QDs in WSe$
_2$~\cite{SrivastavaNNano2015,KoperskiNNano2015,ChakrabortyNNano2015,HeNNano2015, TonndorfOptica2015}. We note that the ratio of the intensity of phonon replica to the parent peak, which quantifies the strength of the exciton-phonon coupling and is called the Huang Rhys factor, varies from $S$ = 0.2 to 0.9 (see Supplementary Information) and is more than an order of magnitude larger than GaAs based QDs~\cite{HeitzPRL1999}.

Having established that we observe strong phonon-replica ($b$-doublets) of the localized neutral exciton ($a$-doublets) with the phonon involved being the $E''$($\Gamma$) CP, we analyze the polarization properties of the doublets. Fig.~3a shows that the red and blue peaks of the D3a parent peak are linearly polarized and are orthogonal to each other. This is consistent with D3a doublet's assignment as a localized $X^0$. Similar linearly polarized emission is observed for other $a$-peaks too (see Supplementary Information). However, the $b$-peaks, in spite of being phonon-replica and inheriting all their properties from the parent $a$-peaks, show completely unpolarized emission in both linear and circular basis measurements (Fig.~3b $\&$ 3c). This stark difference in the polarization property of the phonon replica with respect to the parent peak is particularly puzzling given the coherent nature of the Raman-like phonon emission. 

To explain this anomaly, we first recall that the phonons involved in the scattering process giving rise to the replicas are doubly-degenerate, orthogonal CPs with $l_{\mathrm{phonon}}$ = $\pm$ 1. Moreover, the linearly polarized localized $X^0$ can be thought of as a superposition of $\sigma^+$ and $\sigma^-$ photons and we have a situation identical to that illustrated in Fig.~1d. As a consequence, the combined state of the emitted phonon and replica photon system after the scattering is maximally entangled. As we only measure the polarization of the replica photon and do not have access to the information of phonon AM, we must trace out the phonon subsystem in order to find the state of the photon after it is measured. Starting with the maximally entangled state of Eq.~1 and making use of the fact that the two phonon states with opposite AM are orthogonal, we obtain for the photon subsystem,
\begin{eqnarray}
|\psi_{\mathrm{photon}}\rangle = \frac{1}{2}\left( |\sigma^+ \rangle \langle \sigma^+ | + |\sigma^- \rangle \langle \sigma^- | \right),
\end{eqnarray}   
which is a completely mixed state of polarization, independent of the basis. 

In other words, the information of the polarization of the parent peak photon is lost with the phonon which we never measure. Owing to entanglement, the measurement of the phonon AM to be, say $l$ = +1 ($l$ = -1), would simultaneously project the polarization of the replica photon to $\sigma^+$ ($\sigma^-$) state. However, such a measurement of phonon AM is challenging and is left for future experiments. In reality, the emitted CP have finite lifetime of tens of picosecond after which it decays into other lower energy phonon modes due to anharmonicity of the lattice potential~\cite{CaiPRB2014}. This decay can be thought of as a measurement of the phonon subsystem of the entangled state but as the outcome of this measurement is irreversibly lost to the lattice and environment, the photon subsystem is projected to a completely unpolarized state.

Fig.~3d-e shows quantitative comparison of the measured photon polarization to a completely mixed state of polarization. By extracting Stokes parameter from the polarization measurements we reconstruct the density matrix of the photon polarization and estimate a fidelity of $\mathcal{F}$ = 99 $\pm$ 1.5 $\%$ to the completely mixed state. Other QDs exhibit similar polarization behavior and high values of fidelity (see Supplementary Information).

The entanglement scheme described above crucially relies on two set of degeneracies - that of the two $\sigma$-polarized states entering the superposition to give linearly polarized light and that of orthogonal CPs. These degeneracies lead to two indistinguishable paths for the scattering to take place which have definite phase relationship in their probability amplitudes and hence result in the entangled state. Consequently, if any or both of these degeneracies are lifted, one expects the paths to be become distinguishable and the entanglement to be destroyed. While both the degeneracies are protected by time-reversal symmetry, it is difficult to break the CP degeneracy as phonons do not couple well to $B$ in non-magnetic materials like WSe$_2$. However, the degeneracy of the $\sigma$-polarized states can be broken by an out-of-plane $B$ in a valley Zeeman-like effect~\cite{SrivastavaNPhys2015,AivazianNPhys2015}. 

Fig.~4a shows the $B$ dependence of the polarization of the parent and the replica peak analyzed in circular basis. As $B$ increases, the e-h exchange in $X^0$ doublet is overcome and it becomes $\sigma$-polarized. We observe that the replica peak recovers its polarization and becomes $\sigma$-polarized with increasing $B$, closely following the behavior of the parent peak (Fig.~4b). The recovery of polarization is consistent with the entanglement scheme described above and rules out other mechanisms such as the phonon polarization being oriented in arbitrary directions during the emission events. As the phonon emission process is not expected to be affected by the weak $B$ used in our experiments, such a mechanism would imply unpolarized photon emission even under magnetic fields, contrary to our observations. 

The AM selection rules of Fig.~1d predicts a reversal in helicity of the phonon replica with respect to the parent peak. In fact, we do observe a reversal in helicity for the $E''(\Gamma$) mode in non-resonant Raman scattering measurements (see Supplementary Information). However, the polarization of the phonon replica is to a large degree co-polarized with that of the parent peak at finite $B$ (Fig.~4). Although we lack understanding of this behavior and further studies are needed to address it, similar behavior has been reported by several groups for Raman scattering involving CPs. It was found that the helicities of the incident photon and the Raman scattered photon are reversed with respect to each other only when the incident photon energy is far detuned from the free exciton resonance whereas they are co-polarized for a quasi-resonant excitation~\cite{ChenNanoLett2015, DrapchoPRB2017, YoshikawaPRB2017}. In the case of phonon replica of the QD emission, the role of laser is played by parent peak photon while the Raman scattered photon is the replica photon and the relevant resonance is the $X^0$ QD transition. Thus, for the phonon replica process, we are always ``on-resonance" with respect to the relevant resonance of QD transition and our situation is closer to the quasi-resonant case of Raman scattering, consistent with previous studies. 

Our findings lay the groundwork for the realization of quantum-optomechanical platforms in van der Waals materials~\cite{GeimNature2013}. The strong single exciton-phonon coupling in QDs of two-dimensional materials shown here can serve as a source of single, chiral phonons. Future studies can exploit this chiral coupling to manipulate the quantum state of the collective excitation by an all-optical control of the QD or couple two different QDs via a single chiral phonon mode. Furthermore, the chirality of a macroscopic mode is an intriguing possibility which holds potential for engineering non-reciprocal interactions at the quantum level.
\\
\\
\textbf{Methods}
\\
\textbf{Sample fabrication.} Monolayer WSe$_2$ is mechanically exfoliated from the bulk WSe$_2$ crystal (HQ graphene) on polydimethylsiloxane (PDMS). Similarly thin hexagonal boron nitride (hBN) flake (hBN crystal from HQ graphene) is exfoliated on a degenerately doped Si (Si$^{++}$) substrate with 285~nm SiO$_2$ on top. The fabrication of WSe$_2$$/$BN$/$SiO$_2$ stack is done via the PDMS based dry transfer method~\cite{Castellanos-Gomez2DMaterials2014}. Electron beam lithography is used to deposit 30~nm Pd$/$80~nm Au metal contacts on WSe$_2$, which act as source and drain electrodes. The charge carrier density in WSe$_2$ is controlled by applying voltage (Keithley 2400 sourcemeter) to the Si$^{++}$ substrate, with the 285~nm SiO$_2$ acting as the gate dielectric.  
\\
\textbf{PL spectroscopy.} The sample is loaded into a closed-cycle cryostat (BlueFors) equipped with magnetic field ranging from -8 to +8~T and cooled down to $\sim$ 3.5~K. A piezo controller (Attocube systems) is used to position the sample. Photoluminescence spectroscopy was performed using a home-built confocal microscope set-up. The emission was collected using an aspheric lens (0.55 NA) and directed to a high-resolution (focal length$:$ 750 mm) spectrometer where it was dispersed by a 1200 g$/$mm or 300 g$/$mm grating (both blazed at 750~nm). A liquid nitrogen-cooled charge coupled device (Princeton Instruments SP-2750, PyLoN 1340 $\times$ 400 pixels CCD) was used as detector. Two excitation sources are used$:$ a HeNe laser at 632.8~nm with power of 4~$\mu$W or a mode-hop-free tunable continuous-wave Ti:Sapphire laser (M Squared) with resolution of 0.1~pm and power of 300~nW. The spot size for the Ti:Sapphire laser and the HeNe laser is $\sim$1~$\mu$m and $\sim$2-5 $\mu$m, respectively. Wavelength of the Ti:Sapphire laser is tuned for resonance to the localized excitons to increase the QD emission intensity. Polarization of the incident laser is controlled using a polarizer together with a $\lambda$$/$2 or a $\lambda$$/$4 plate. Polarization measurements were performed by using a Wollaston prism which separates light into $s$- and $p$-components. A $\lambda$$/$4 plate is placed after the Wollaston prism to convert the linearly polarized light into circularly polarized signal, so that the signal will be insensitive to the grating efficiency. A $\lambda$$/$2 (super-achromatic 600-2700~nm, Thorlabs) or $\lambda$$/$4 (zero order @ 780~nm) plate is placed before the Wollaston prism for the linear or circular basis measurement. The influence of blinking on polarization measurements is eliminated in this setup (see Supplementary Information). In all the magnetic field dependence measurements, $B$ is applied perpendicular to the plane of the sample.
\\
\textbf{Raman spectroscopy} Raman scattering measurements were carried out at room temperature. The excitation source is a He-Cd laser at 441.6 nm with power of 0.5 mW. The backscattered signal was collected through a 100x objective, directed to a micro-Raman spectrometer (Horiba-HR Evo) where it was dispersed by a 1800 g/mm grating and finally detected using a liquid nitrogen cooled charge-coupled device with a spectral resolution of $\sim$ 1 cm$^{-1}$.
\\
\textbf{Phonon dispersion calculation.} We use Quantum Espresso Code to calculate the phonon dispersion of monolayer WSe$_2$. The norm-conserving pseudopotential within the local density approximation (LDA) of Perdew-Zunger is used. The kinetic energy Cutoff for charge density is 260 Ry. The first Brillouin zone is sampled with a 31 $\times$ 31$\times$ 1 Monkhorst-Pack grid. The vacuum region thickness is 20 \AA. The optimized equilibrium lattice constant of monolayer WSe$_2$ is 3.13 \AA.

\vspace{1 cm}

\textbf{Acknowledgments} We acknowledge many enlightening discussions with Ata\c{c} Imamo\u{g}lu, Weibo Gao  and Martin Kroner. We also acknowledge technical help from Timothy Neal and Eva Liu. A. S. acknowledges support from Emory University startup funds and NSF through the EFRI program-grant \# EFMA-1741691. L.Z. thanks M. Gao for helpful calculation and discussion and acknowledges support from the National Natural Science Foundation of China (grant No. 11574154). Q.X. gratefully acknowledges strong support from Singapore National Research Foundation via NRF-ANR joint grant (NRF2017-NRF-ANR002 2D-Chiral) and Singapore Ministry of Education via AcRF Tier2 grant (MOE2017-T2-1-040) and Tier1 grants (RG 113/16 and RG 194/17)\\

\textbf{Author Contributions} ~$^\dagger$ $\textbf{X. C., X. L.}$ and $\textbf{S. D.}$ contributed equally to this work. X. C., X. L., S. D. and Q. Y. carried out the quantum dot measurements and S. L. measured the Raman data. X. L. and X. W. prepared the samples. A. S., L. Z. and Q. X. supervised the project. All authors were involved in analysis of the experimental data and contributed extensively to this work.\\

\textbf{Author Information} The authors declare that they have no
competing financial interests. $^*$Correspondence and requests for
materials should be addressed to A.S. (ajit.srivastava@emory.edu).

\textbf{Competing financial interests} The authors declare no competing financial interests.

\newpage

{\bf Figure 1: Chiral phonons and phonon-photon entanglement}  {\bf a,} Calculated phonon dispersion of monolayer WSe$_2$ along the $\Gamma$-$K$ direction of the Brillouin zone. Chiral phonons (CPs) with (pseudo-)angular momentum (AM) $l=\pm1$ occur at zone center ($\Gamma$-point) and zone boundary ($\pm K$-point). TOP: Vibrational normal modes of the doubly-degenerate $E'' (\Gamma)$ mode in monolayer WSe$_2$. While $W$ atoms remains stationary, the vibration of $Se$ atoms can be chosen in any in-plane directions. Superposition of two orthogonal linear vibrations results in chiral phonons with clockwise or counterclockwise motion. {\bf b,} Raman spectra of  $E''$($\Gamma$) mode in WSe$_2$ of different thickness. The energy of $E''$($\Gamma$) is measured to be 176 cm$^{-1}$ (21.8 meV). The peak is forbidden in monolayer due to symmetry constraints. {\bf c,d} Schematic of phonon-photon entanglement. The circularly polarized states ($\sigma^+$/$\sigma^-$) with AM of $ l$ = +1/-1 are degenerate in WSe$_2$ due to time-reversal symmetry.  The $ l$ = +1 (-1) phonon can only couple a $\sigma^-$($\sigma^+$) photon to a scattered photon with angular momentum  of $l$ = +1 (-1) due to conservation of AM. The indistinguishability of the two paths in this ``which-way" scattering process leads to entanglement of phonon-photon.
\\

{\bf Figure 2: QDs and their phonon replicas in monolayer WSe$_2$} {\bf a,} PL intensity map as a function of back-gate voltage ($ V_\mathrm{g}$). The attractive Fermi-polaron peak ($X_{\mathrm{ap}}$) and the free exciton $X^0$ are identified. Defect peak ({\textit D}) is observed when -90 V $< V_\mathrm{g} <$ 20 V, and concurrently sharp QD-like peaks appear with energy lower than {\textit D}. {\bf b,} Gate-dependent PL intensity map of QDs. The assignment of a QD group is based on correlated spectral jittering pattern (Highlighted in  {\bf b}: green stars for D4 group; red triangles and squares for D3 group). Each group contains two doublets, with the high- and low-energy doublets named as {\textit a} and  {\textit b} ({\textit e.g.} D3a and D3b), respectively. Dashed line indicates the onset V$_\mathrm{g}$ for each group. {\bf c,} A cross-sectional PL spectrum at $V_\mathrm{g}$ = -78 V. The splitting energy of  {\textit a} doublets is identical to that of the corresponding {\textit b} doublets. The energy spacing between the {\textit a} and {\textit b} doublets is 21.8 meV, which is consistent with the energy of $E''$($\Gamma$) phonon. Inset shows similar behavior for QD D6. {\bf d,} Power dependence plot of the D3 group. The lines are the power-law fitting $I \propto P^\alpha$. The extracted values of $\alpha$ are same within the standard deviation for the four peaks in D3 group. Sublinear dependence further confirms the sharp peaks are related to trapped excitons. {\bf e,} Zeeman splitting between the two peaks in the D3a (top) and D3b (bottom) doublets as a function of $B$ field. The g-factors are same for both doublets within the error bar. {\bf d} and {\bf e} were measured at $V_\mathrm{g}$ = -80 V. 
\\

{\bf Figure 3: Polarization dependence of QDs and their phonon replicas in monolayer WSe$_2$.} {\bf a,} Polarization of D3a doublet measured in the linear basis. The lines are sin$^2\theta$/cos$^2\theta$ fits to the experimental data (dots), showing the two peaks are cross-polarized. {\bf b, c,} Polarization of the red peak from the D3b doublet in the linear (b) and circular (c) bases. Black lines are the average values, and the brown shaded regions represent the standard deviation of the experimental data (dots). Green dashed line in (b) shows an example of the linearly polarized emission in linear basis measurement. The red dashed circle with radius of 0.5 can be either circularly polarized emission or an unpolarized light source. Further measurements in circular basis (c) distinguish unpolarized emission from circular polarization. Orange dash line in (c) shows an example of circularly polarized emission in circular basis measurement while the red dashed circle with radius of 0.5 represents unpolarized emission. {\bf d, e,} Real (d) and imaginary (e) parts of the density matrix for the polarization state of D3b red peak. The hollow caps in (d) indicate the values of diagonal matrix elements (0.5) for a completely mixed state. Polarization measurements were carried out at $V_\mathrm{g}$ = -80 V. 
\\

{\bf Figure 4: Recovery of polarization of phonon replicas in magnetic field.} {\bf a,} PL intensity map of D2a and D2b doublets as a function of $B$ field measured in the circular basis. As out-of-plane $B$ field increases, the splitting between the two peaks comprising the doublets $a$ and $b$ increases. Efficient thermalization results in only the low energy peak being visible in PL. In addition, the intensity of D2a red peak increases in the $\sigma^+$ configuration, while decreases in the $\sigma^-$ detection (left). D2b red peak shows the same trend of behavior (right). {\bf b,} Circular dichroism (CD) of D3a blue peak and D3b blue peak as a function of $B$ field. The circular polarization of D3b is smaller than D3a from -0.8 T to 1 T. Under $B$ field, CD of D3b peak recovers from zero and follows that of D3a peak. $B$ field measurements were carried out at $V_\mathrm{g}$ = -80 V. 
\\

\begin{figure}
\includegraphics[scale=1.5]{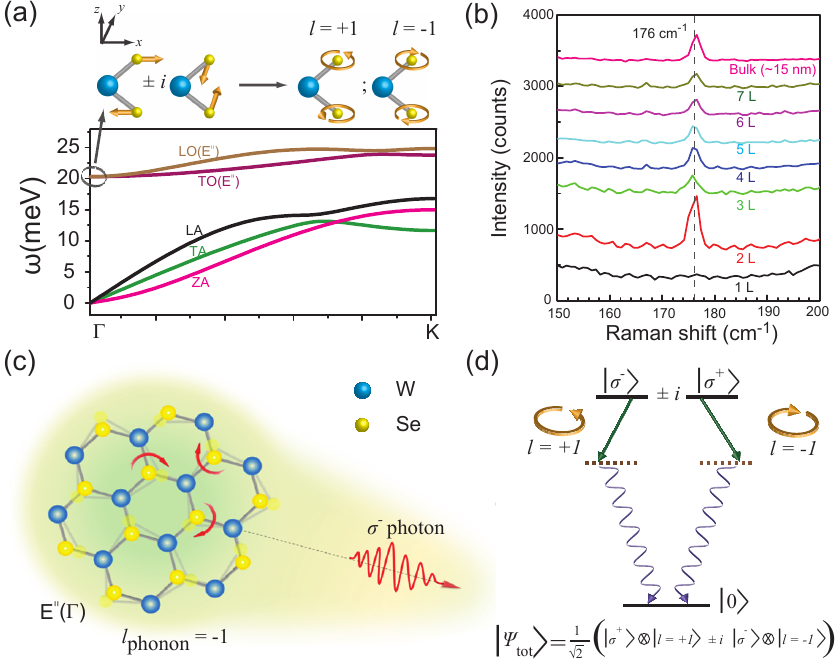}
\end{figure}

\clearpage

\begin{figure}
\includegraphics[scale=0.6]{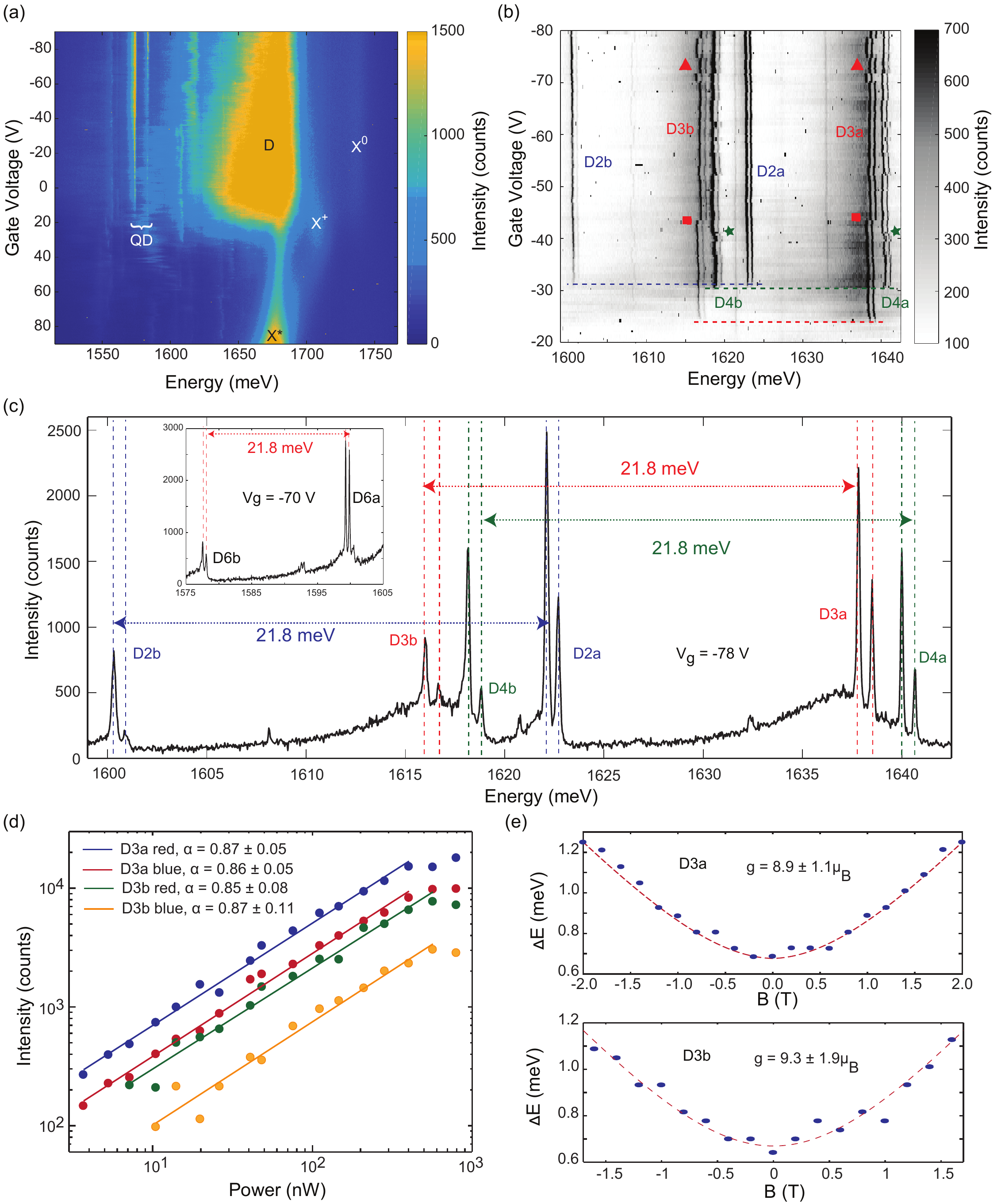}
\end{figure}

\clearpage

\begin{figure}
\includegraphics[scale=0.7]{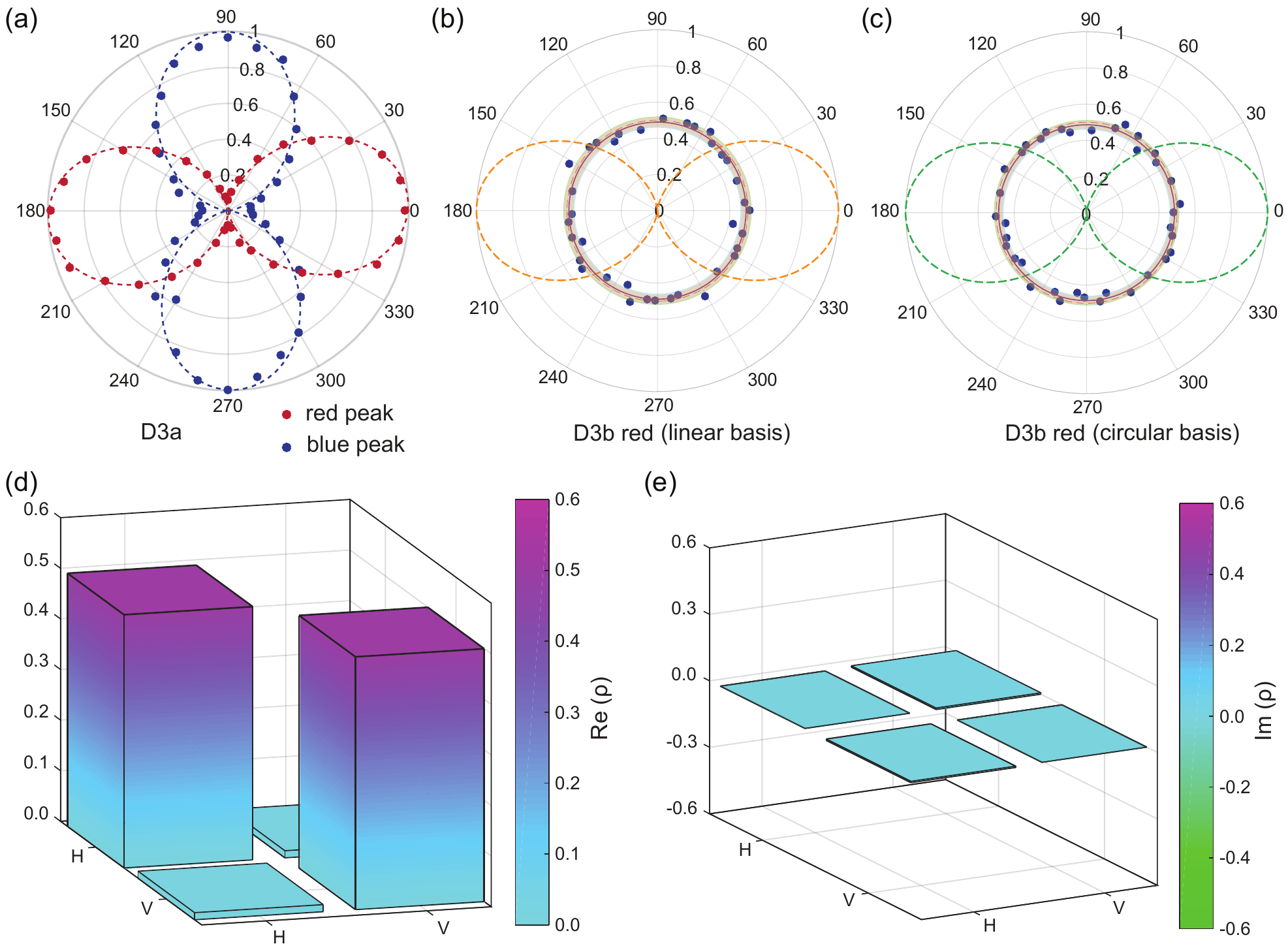}
\end{figure}

\clearpage

\begin{figure}
\includegraphics[scale=0.9]{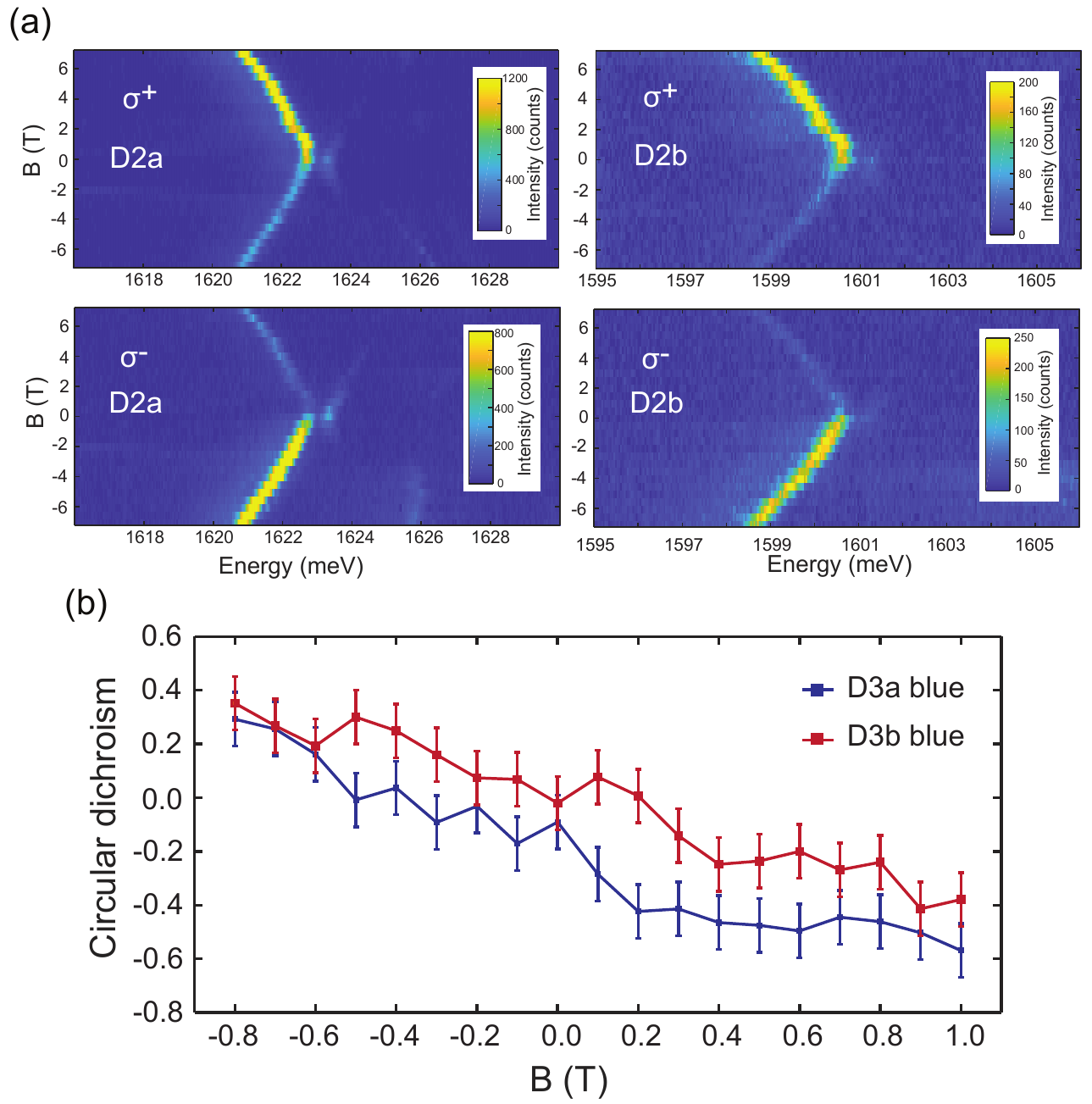}
\end{figure}

\clearpage

%
%
%
%
%


\textbf{Supplementary Information:}
\title{Entanglement of single-photons and chiral phonons in atomically thin WSe$_2$}

\author{Xiaotong Chen$^{1*}$}
\author{Xin Lu$^{1*}$}
\author{Sudipta Dubey$^{1*}$}
\author{Qiang Yao$^{1}$}
\author{Sheng Liu$^{2}$}
\author{Xingzhi Wang$^{2}$}
\author{Qihua Xiong$^{2,3}$}
\author{Lifa Zhang$^{4}$}
\author{Ajit Srivastava$^{1	\dagger}$}
\affiliation{$^1$Department of Physics, Emory University, Atlanta 30322, Georgia, USA}
\affiliation{$^2$Division of Physics and Applied Physics, School of Physical and Mathematical Sciences, Nanyang Technological University, Singapore 637371, Singapore}
\affiliation{$^3$NOVITAS, Nanoelectronics Centre of Excellence, School of Electrical and Electronic Engineering, Nanyang Technological University, Singapore, 639798, Singapore and}
\affiliation{$^4$School of Physics and Technology, Nanjing Normal University, No.1 Wenyuan Road Qixia District, Nanjing, P.R.China 210046.}

\maketitle


\textbf{\underline{Supplementary Information Contents:}}\\
1. Optical characterization of the monolayer WSe$_2$ field effect transistor (FET) device.\\
2. Gate-dependent photoluminescence (PL) of quantum dots (QDs) and their phonon replicas.\\
3. Polarization measurements to mitigate the effect of blinking.\\
4. Polarization state and fidelity of phonon replica.\\
5. Symmetry analysis of $E''$ phonon mode in monolayer WSe$_2$.\\
6. Photoluminescence excitation spectroscopy (PLE) of D2a, D3a and D4a doublets.

\clearpage
\textbf{1. Optical characterization of the monolayer WSe$_2$ field effect transistor (FET) device}\\

We use optical contrast and photoluminescence (PL) spectroscopy for thickness identification. Monolayer WSe$_2$ has strong emission at room temperature due to the direct-gap transition at $\pm K$-point. As shown in Figure S1, a prominent and asymmetric peak centered at $\sim$750 nm is observed from monolayer WSe$_2$.  As thickness increases, WSe$_2$ evolves to be an indirect-gap semiconductor with much weaker PL intensity. The emission energy also red-shifts in thicker layers, and the emission from bilayer is centered at $\sim$800 nm \cite{Tonndorf13}.

\begin{figure}[H]
\begin{center}
\includegraphics[width=120mm]{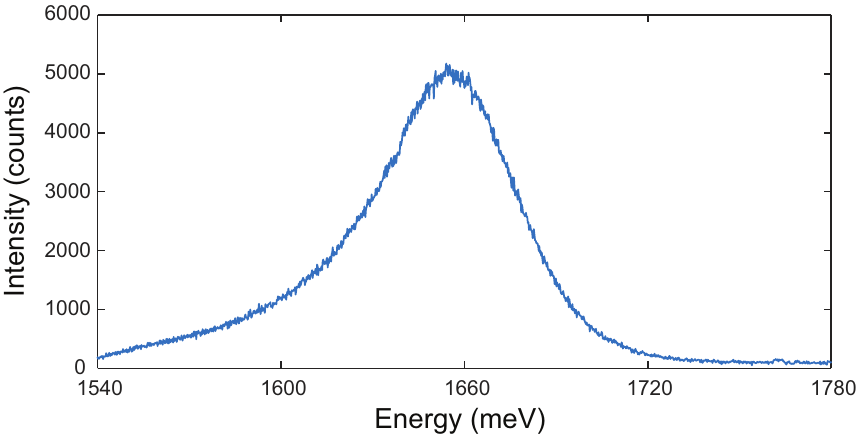}\\
\end{center}
{\bf Figure S1: PL of monolayer WSe$_2$ at room temperature.}
A prominent and asymmetric peak centered at 1654 meV ($\sim$750 nm) is detected with $\lambda_{\mathrm{exc}}$ = 632.8 nm and $P$ = 16 $\mu$W, confirming the flake to be monolayer. 
\end{figure}

Figure S2a shows the optical image of monolayer WSe$_2$/BN stack on a degenerately doped Si (Si$^{++}$) substrate with 285 nm SiO$_2$ on top. The charge carrier density in WSe$_2$ is controlled by applying voltage to the Si$^{++}$ substrate, with the 285 nm SiO$_2$ acting as the gate dielectric. At negative gate voltage ($V_\mathrm{g}$) when the sample is depleted of electrons or hole-doped, we observed six groups of quantum dots (QDs) and their phonon replicas. Figure S2b-S2d display the spatial map of QDs in the monolayer WSe$_2$ FET device. 

As the six QD groups all appear on WSe$_2$/BN stack , it is important to trace the origins of the QDs. We thus performed a gate dependent PL mapping. Figure S3 demonstrates that the defect peak ($D$) from WSe$_2$ and QDs appear at almost the same $V_\mathrm{g}$, which implies that the QDs originate from WSe$_2$ and not from hBN spacer layer.

\begin{figure}[H]
\begin{center}
\includegraphics[width=160mm]{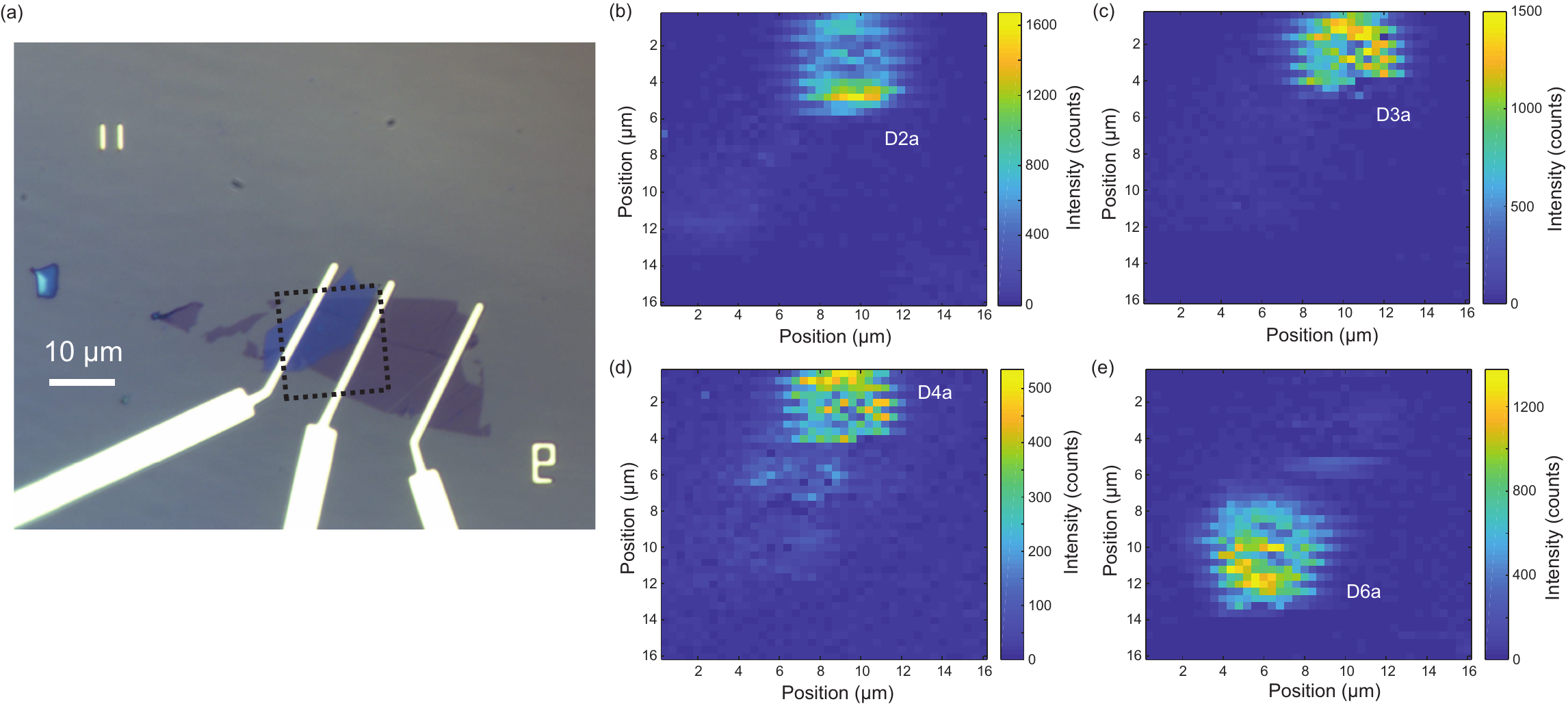}
\end{center}
{\bf Figure S2: Spatial location of the quantum dots in monolayer WSe$_2$ FET device.}  {\bf a,} Optical image of the monolayer WSe$_2$/BN stack under white light illumination. The dashed square indicates the scanning area in the PL mapping measurements. {\bf b-e,} PL intensity map of QDs centered at 1662.4 meV ({\bf b}, D2a red peak), 1638.2 meV ({\bf c}, D3a red peak), 1640.2 meV ({\bf d}, D4a red peak) and 1599.3 meV ({\bf e}, D6a red peak) over a 16~$\mu$m $\times$ 16~$\mu$m scanning area. A back gate voltage, $V_\mathrm{g}$ = -80 $V$ was applied during the measurement. Excitation wavelength, $\lambda_{\mathrm{exc}}$ = 732.5 nm and incident power, $P$ = 300 nW.
\end{figure}

\begin{figure}[H]
\begin{center}
\includegraphics[width=90mm]{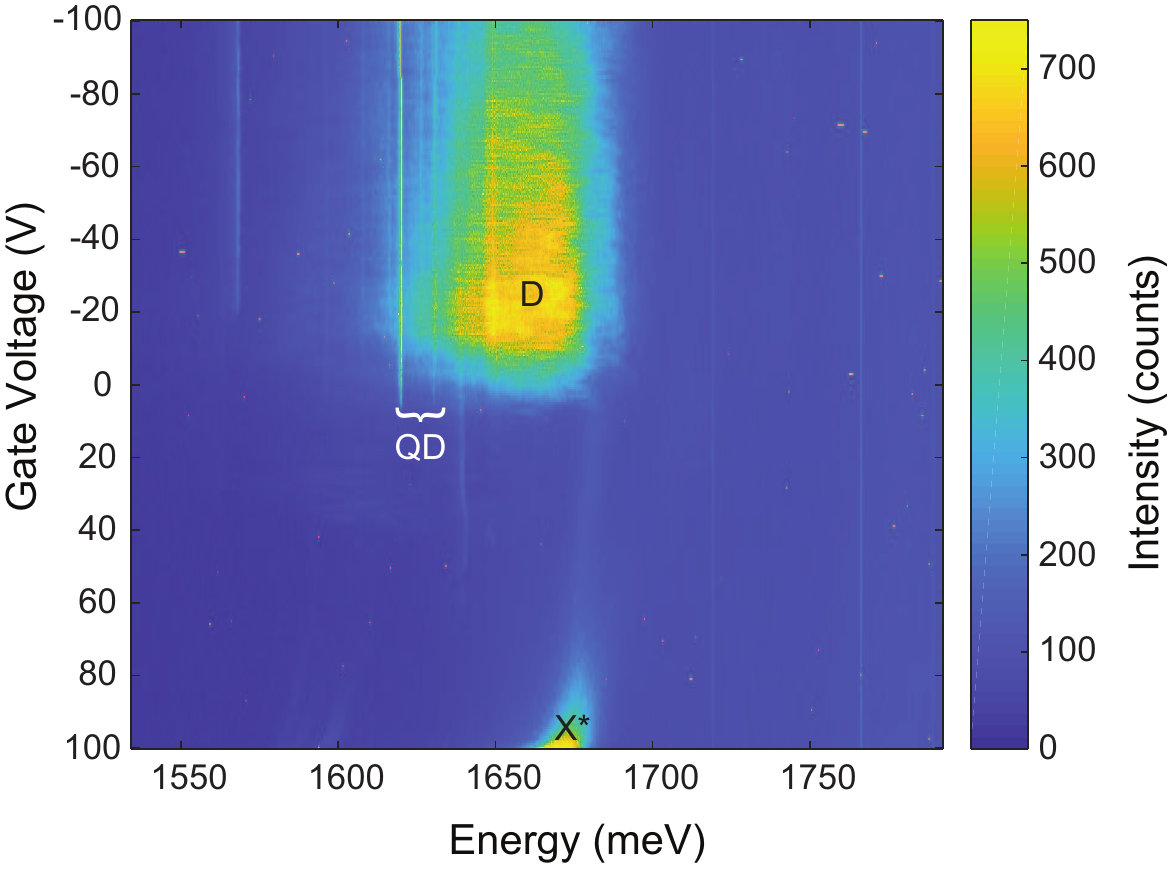}\\
\end{center}
{\bf Figure S3: Gate-dependent PL intensity map of defect peak and QDs.} Gate dependence of PL intensity at a different region compared to Fig.~2a of the main text. As $V_\mathrm{g}$ changes from 100 to -100 V, the defect peak ($D$) from WSe$_2$ and QDs appear at almost the same $V_\mathrm{g}$, which shows that the QDs originate from WSe$_2$ and not from hBN spacer layer. $\lambda_{\mathrm{exc}}$ = 632.8 nm and $P$ = 1 $\mu$W.\\
\end{figure}

\clearpage
{\bf 2. Gate-dependent PL of QDs and their phonon replicas.}

We have shown gate-dependent PL intensity map of D2, D3 and D4 QD groups in the main text (Figure 2b). In Figure S4, we present the PL mapping of D1, D5 and D6 QD groups as a function of $V_\mathrm{g}$. The assignment of a QD group is based on correlated spectral jittering pattern, as highlighted by solid symbols. Each group contains two doublets, with the high- and low-energy doublets named as {\it a} and {\it b} ({\it e.g.} D1a and D1b), respectively. The splitting energy of {\it a} doublets is identical to that of the corresponding {\it b} doublets. The fine structure splitting for all the doublets are summarized in Table I. The energy spacing between the {\it a} and {\it b} doublets is 21.8 meV, which is equal to the energy of $E''(\Gamma$) phonon in WSe$_2$ (Figure 1b). We thus conclude that the lower energy {\it b}-peaks are the chiral phonon-replica of the parent {\it a}-peaks.

The intensity ratio of the phonon replica (first replica:{\it b}-doublet) to the parent peak (zero-phonon line:{\it a}-doublet), which quantifies the strength of the exciton-phonon coupling, is called the Huang Rhys factor ($S$). The intensity of the $n$-th phonon replica can be expressed as \cite{Huang50}: $I_n=S^ne^{-S}/n!$. Since some of the blue peaks from the {\it b}-doublet is weak ({\it e.g.} D2b and D5b), here we only consider the intensity of the red peak. $I_0$ is the intensity of the red peak from {\it a}-doublet, and $I_1$ is that from the {\it b}-doublet. By using the function $I_1=I_0S$, we obtain the Huang Rhys factor for the six QD groups, as shown in Table II. The parameter $S$ varies from 0.18 to 0.88 in our WSe$_2$ QDs, which is an order of magnitude larger than InAs/GaAs QDs \cite{Heitz99}. But we did not observe the second phonon replica in our QDs, even in the D4 group having a large Huang Rhys factor (Figure 2).

\begin{figure}[H]
\includegraphics[width=160mm]{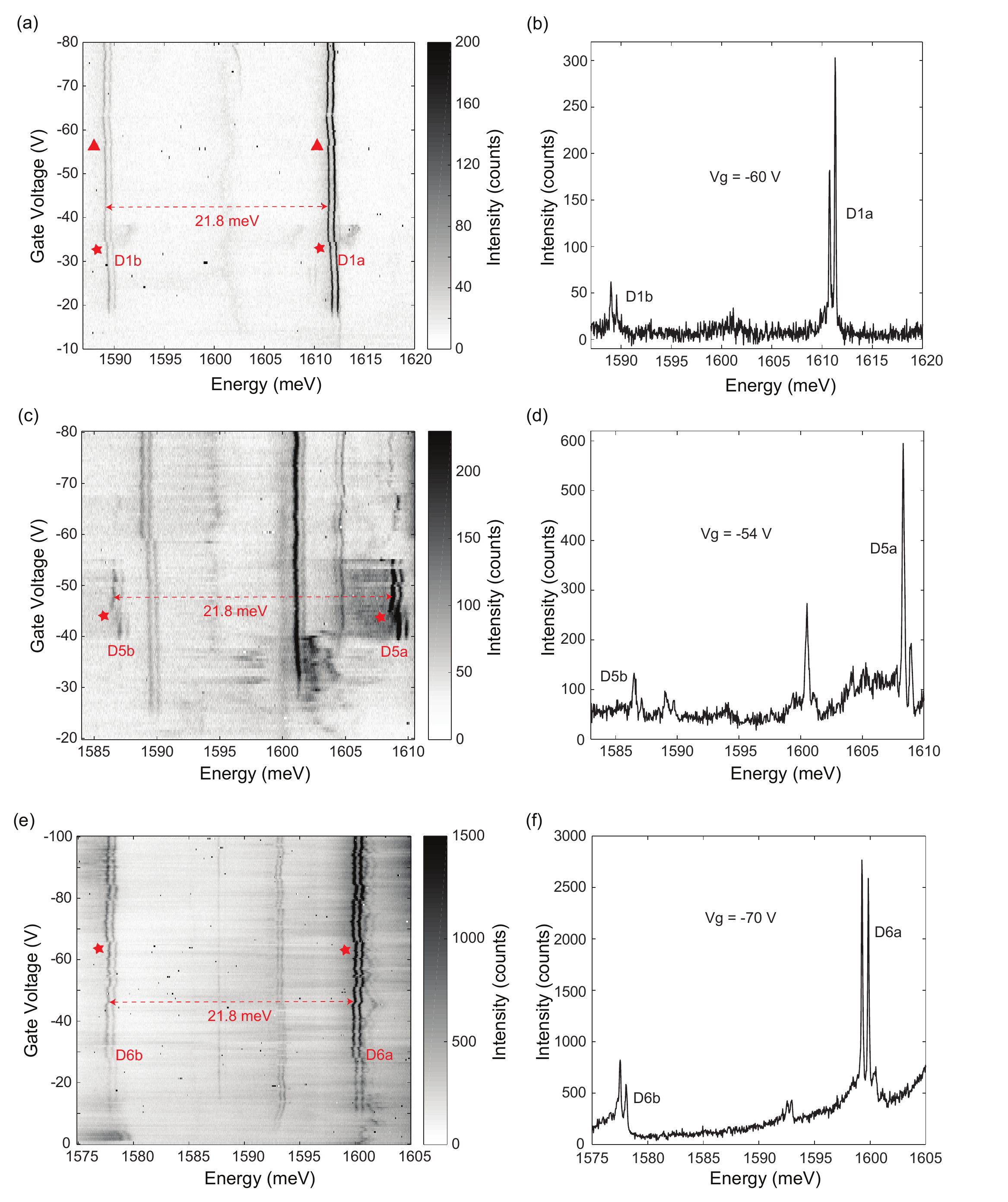}
{\bf Figure S4: Gate-dependent PL intensity map of D5 and D6 groups.}   {\bf a,} PL intensity map of D1 group as a function of $V_\mathrm{g}$. $\lambda_{\mathrm{exc}}$ = 735 nm and $P$ = 43 nW. {\bf b,} A cross-sectional PL spectrum at $V_\mathrm{g}$ = -60 V.  {\bf c,} PL intensity map of D5 group as a function of $V_\mathrm{g}$. $\lambda_{\mathrm{exc}}$ = 745 nm and $P$ = 300 nW. {\bf d,} A cross-sectional PL spectrum at $V_\mathrm{g}$ = -54 V.
{\bf e,} PL intensity map of D6 group as a function of $V_\mathrm{g}$. $\lambda_{\mathrm{exc}}$ = 728 nm and $P$ = 300 nW.  {\bf f,} A cross-sectional PL spectrum at $V_\mathrm{g}$ = -70 V. The assignment of a QD group is based on correlated spectral jittering pattern (Highlighted by solid symbols). The energy spacing between the {\it a} and {\it b} doublets is 21.8 meV, consistent with the energy of $E''(\Gamma$) phonon in WSe$_2$. Wavelength of the incident laser is tuned for resonance to the QD group.
\end{figure}

\begin{table}[H]
\caption{Energy splitting of doublets}
\centering 
\begin{tabular}{c c c}
\hline\hline 
Group & doublet \textbf{\textit{a}} & doublet \textbf{\textit{b}} \\ 
\hline
D1 & 0.58$\pm$0.04 & 0.56$\pm$0.08  \\
D2 & 0.57$\pm$0.06 & 0.57$\pm$0.06  \\
D3 & 0.66$\pm$0.02 & 0.64$\pm$0.06  \\
D4 & 0.69$\pm$0.01 & 0.68$\pm$0.06  \\
D5 & 0.61$\pm$0.04 & 0.58$\pm$0.10  \\
D6 & 0.57$\pm$0.05 & 0.56$\pm$0.05  \\
\hline
\end{tabular}
\end{table}

\begin{table}[H]
\caption{Huang Rhys factors of the QDs in WSe$_2$}
\centering 
\begin{tabular}{c c}
\hline\hline 
Group & Huang Rhys factor\\
\hline
D1 & 0.28$\pm$0.08\\
D2 & 0.32$\pm$0.16  \\
D3 & 0.27$\pm$0.04  \\
D4 & 0.88$\pm$0.17  \\
D5 & 0.18$\pm$0.05\\
D6 & 0.26$\pm$0.05\\
\hline
\end{tabular}
\end{table}

\clearpage
Excitation power-dependence measurements further confirm that the {\it b}-peaks are phonon replicas and not other excitonic complexes. Figure S5 demonstrates that all the peaks in D2 group exhibit a sub-linear, power-law behavior with emission intensity ($I$) scaling with the incident power ($P$) as $P^{\alpha}$ with $\alpha\sim0.6$.

\begin{figure}[H]
\vskip 15pt
\begin{center}
\includegraphics[width=70mm]{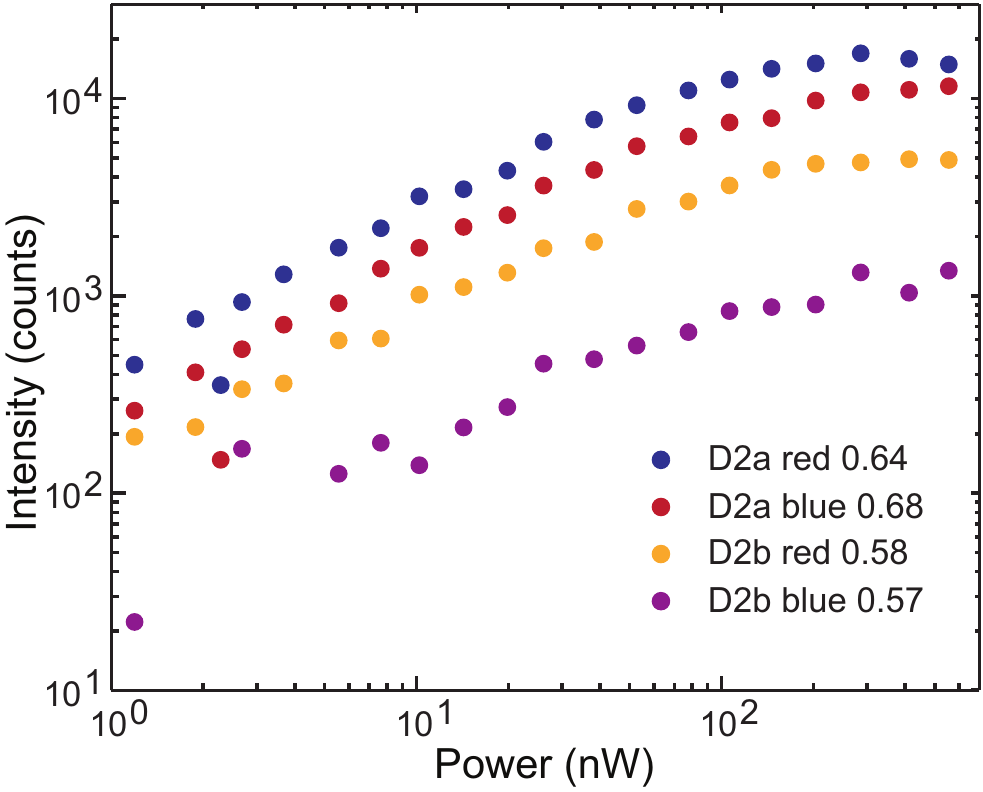}\\
\end{center}
{\bf Figure S5: Power dependence of peaks in D2 group.} The intensity of PL was fitted with power law function \textbf{\textit{$I \propto P^{\alpha}$}}. The extracted values of $\alpha$ indicate that the PL intensity is sublinear with $P$ as is expected for localized emitters. $\alpha$ is the same within error for $a$- and $b$- doublets as is expected for phonon replica. $V_\mathrm{g}$ = -80 $V$ was applied during the measurement. Excitation wavelength, $\lambda_{\mathrm{exc}}$ = 732.5 nm.
\end{figure}

\newpage

Figure S6 shows that both {\it a}- and {\it b}-doublets display Zeeman splitting in an out-of-plane magnetic field ($B$) with almost identical g-factor as is expected for the parent and phonon replica peaks. The measured g-factors for D1, D2, D3 and D5 groups are summarized in Table III.

\begin{figure}[H]
\begin{center}
\includegraphics[height=70mm]{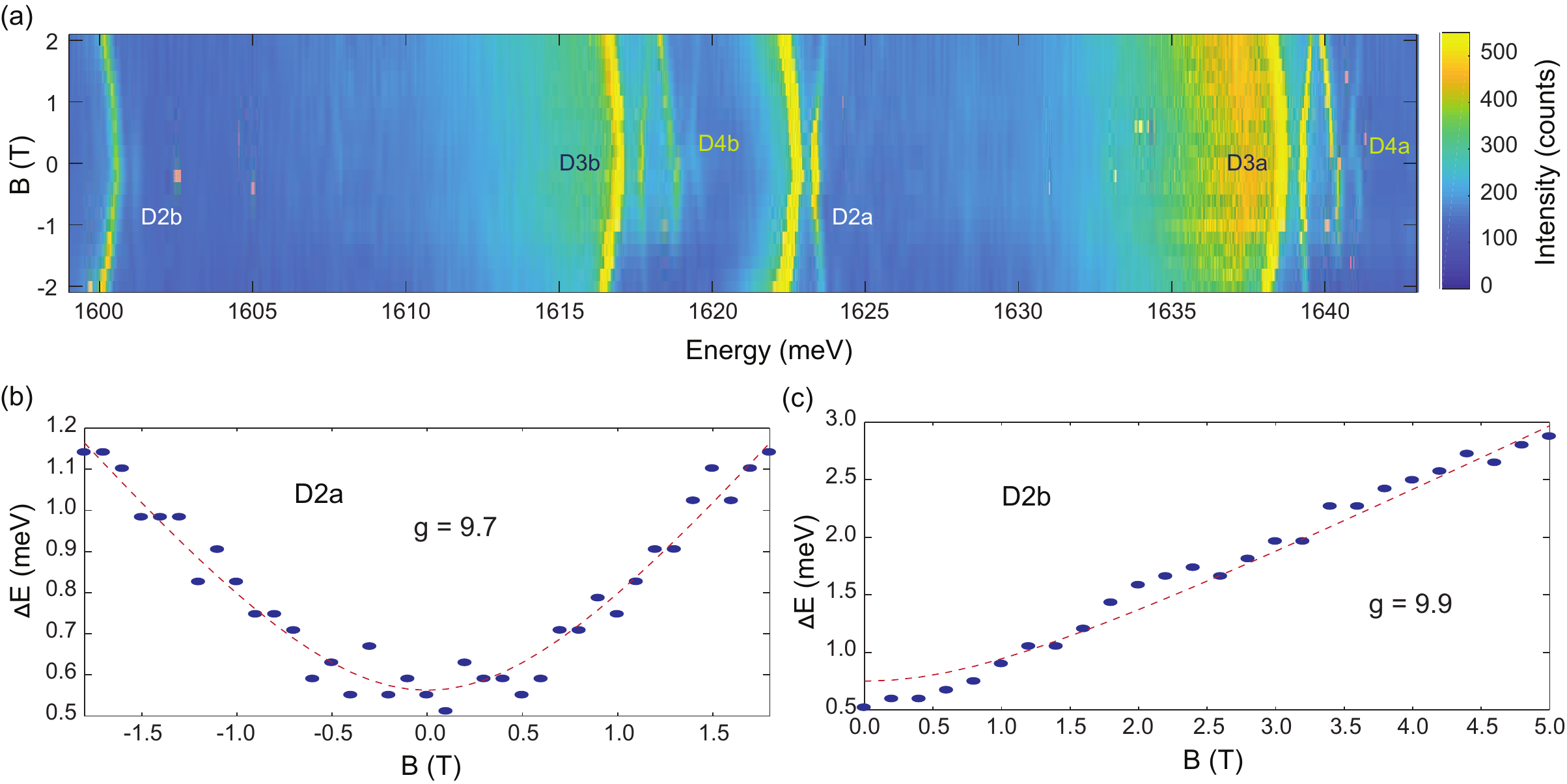}\\
\end{center}
{\bf Figure S6: $B$ field-dependent PL  intensity map and g-factor of D2 group.}  {\bf a,} PL intensity map of D2, D3 and D4 QD groups as a function of $B$ field. Spectra were taken at every 0.2 T with $\lambda_{\mathrm{exc}}$ = 732.5 nm and $P$ = 300 nW. {\bf b,c} Zeeman splitting between the D2a (b) and D2b (c) doublets  as a function of $B$ field. The g-factors are similar for both doublets, as is expected for phonon replica. 
\end{figure}

\begin{table}[H]
\caption{The g-factor of QDs}
\centering 
\begin{tabular}{c c c}
\hline\hline 
Group & doublet \textbf{\textit{a}} & doublet \textbf{\textit{b}} \\ 
\hline
D1 & 9.4$\pm$0.3 & 10.2$\pm$0.6  \\
D2 & 9.7$\pm$0.8 & 9.9$\pm$0.5  \\
D3 & 8.9$\pm$1.1 & 9.3$\pm$1.9  \\
D5 & 9.6$\pm$1.2 & 8.6$\pm$0.9  \\
\hline
\end{tabular}
\end{table}

\newpage

\textbf {3. Polarization measurements to mitigate the effect of blinking.}\\

\begin{figure}[H]
\vskip 10pt
\includegraphics[width=160mm]{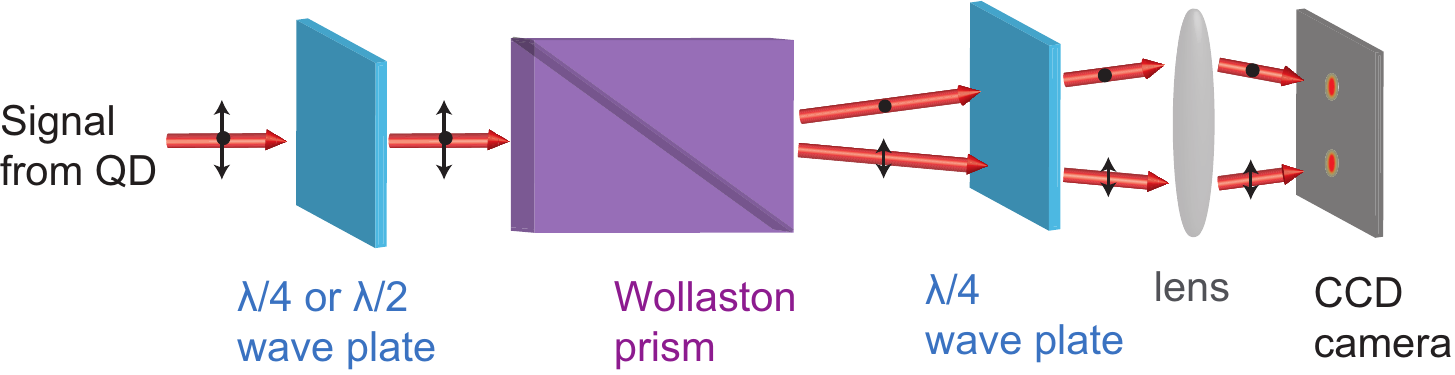}
{\bf Figure S7: Schematic of the setup for polarization measurements to mitigate the effect of blinking.} Polarization measurements were performed by using a Wollaston prism (WP) which separates light into $s$- and $p$-components. A $\lambda$/4 wave-plate is placed after the WP to convert the linearly polarized light into circularly polarized signal, so that the signal will be insensitive to the grating efficiency. A $\lambda$/2 or $\lambda$/4 wave-plate is placed before the WP for the linear or circular basis measurement. The influence of blinking on polarization measurements is eliminated in this setup.
\end{figure}

We use linearly polarized light to excite the QD and the QD's emission is collected through setups for both linear and circular bases. In order to avoid the influence of QD intensity intermittency, a Wollaston prism (WP) is used to analyze the polarization as the two components from WP will be simultaneously measured by the detector.  A $\lambda$/4 wave-plate is placed after the WP to convert the linear light into circularly polarized signal, so that the signal will be insensitive to the grating efficiency, as shown in Figure S7. In linear polarization basis, the setup consists of a $\lambda$/2 wave-plate and a WP. Light passing through the WP will be separated into two orthogonal and linearly polarized components. Their intensities are $  I_{•1}$ and $  I_{•2}$. The linearly polarized component's ratio, $  I_{•1}$/ ($  I_{•1}$+ $  I_{•2}$), in any direction can be obtained by rotating the $\lambda$/2 wave-plate. Figure 3a and Figure S8a show that the D3a and D4a doublets are linear- and cross-polarized. While both circularly polarized emission and an unpolarized light source show a circle with radius of 0.5 in linear basis measurements (Figure 3b and Figure S8b), we further proceed to measurements in circular basis. The circular basis setup consists of a $\lambda$/4 wave-plate and a WP. The ratio of $  I_{•1}$/ ($  I_{•1}$+ $  I_{•2}$) now represents the percentage of left (or right) cricularly polarized component. As shown in Figure S8c, the intensity of D4b red peak doesn't change in circular base, indicating the phonon replica is unpolarized. Even with circular excitation and even in quasi-resonant scenario, {\textit{b}}-doublets are still unpolarized (Figure S9). Figure S10 shows that {\textit{a}}-doublets are not affected by incident polarization as well.\\

\begin{figure}[H]
\includegraphics[width=160mm]{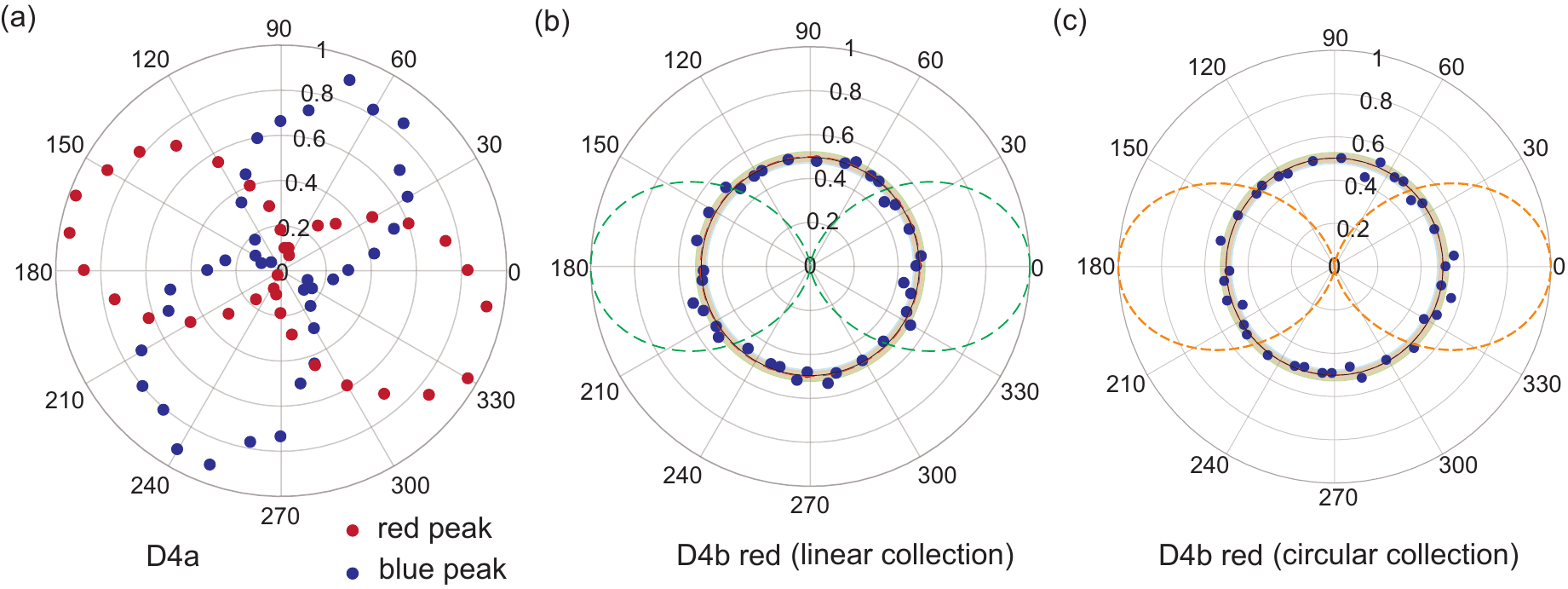} 
{\bf Figure S8: Polarization of the D4 group measured with Wollaston prism.} {\bf a,} Polarization of the D4a doublet measured in the linear basis. The two peaks are cross-polarized. {\bf b,c} Polarization of the red peak from the D4b doublet in the linear (b) and circular (c) bases. Black lines are the average values, and the brown shaded regions represent the standard deviation of the experimental data (dots). (b) Green dashed line shows an example of the linearly polarized emission in linear basis measurement. Red dash circle with radius of 0.5 can be either circularly polarized emission or an unpolarized light source. Further measurements in circular basis distinguish unpolarized emission from circularly polarized emission. Orange dash line in (c) shows an example of circularly polarized emission in circular basis measurement. Red dashed circle with radius of 0.5 in (c) represents unpolarized light source.
\end{figure}

\newpage

\begin{figure}[H]
\begin{center}
\includegraphics[width=90mm]{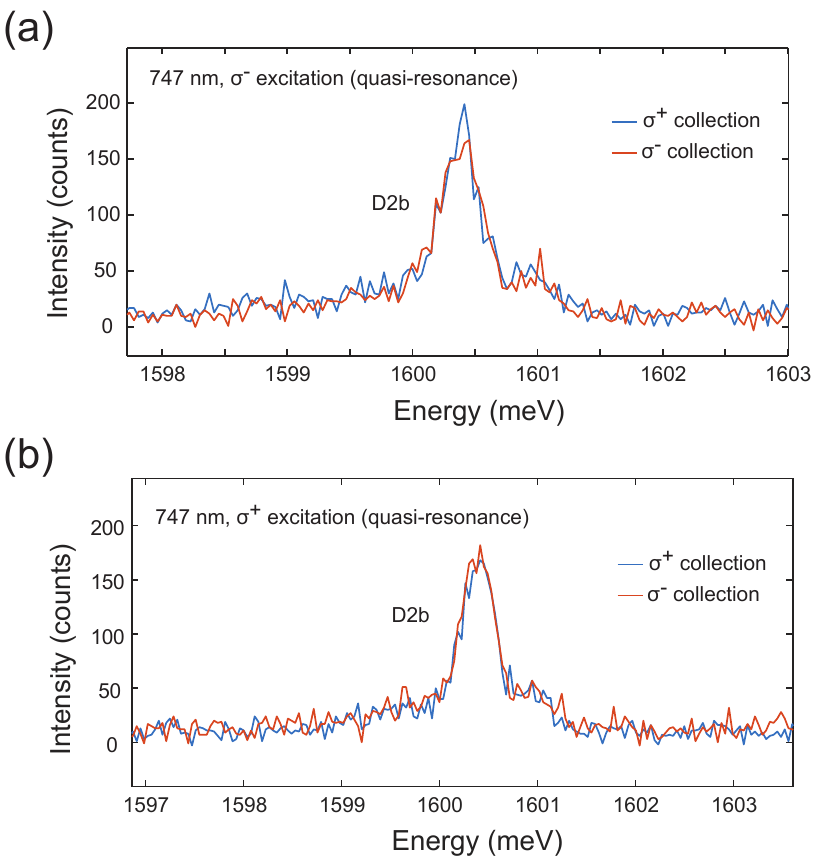} 
\end{center}
{\bf Figure S9: Polarization of D2b doublet measured in the circular basis with quasi-resonant excitation.} {\bf a,} PL Spectrum of D2b doublet with $\sigma^-$ excitation. {\bf b,} PL Spectrum of D2b doublet with $\sigma^+$ excitation. $\lambda_{\mathrm{exc}}$ = 747 nm.
\end{figure}

\begin{figure}[H]
\begin{center}
\includegraphics[width=110mm]{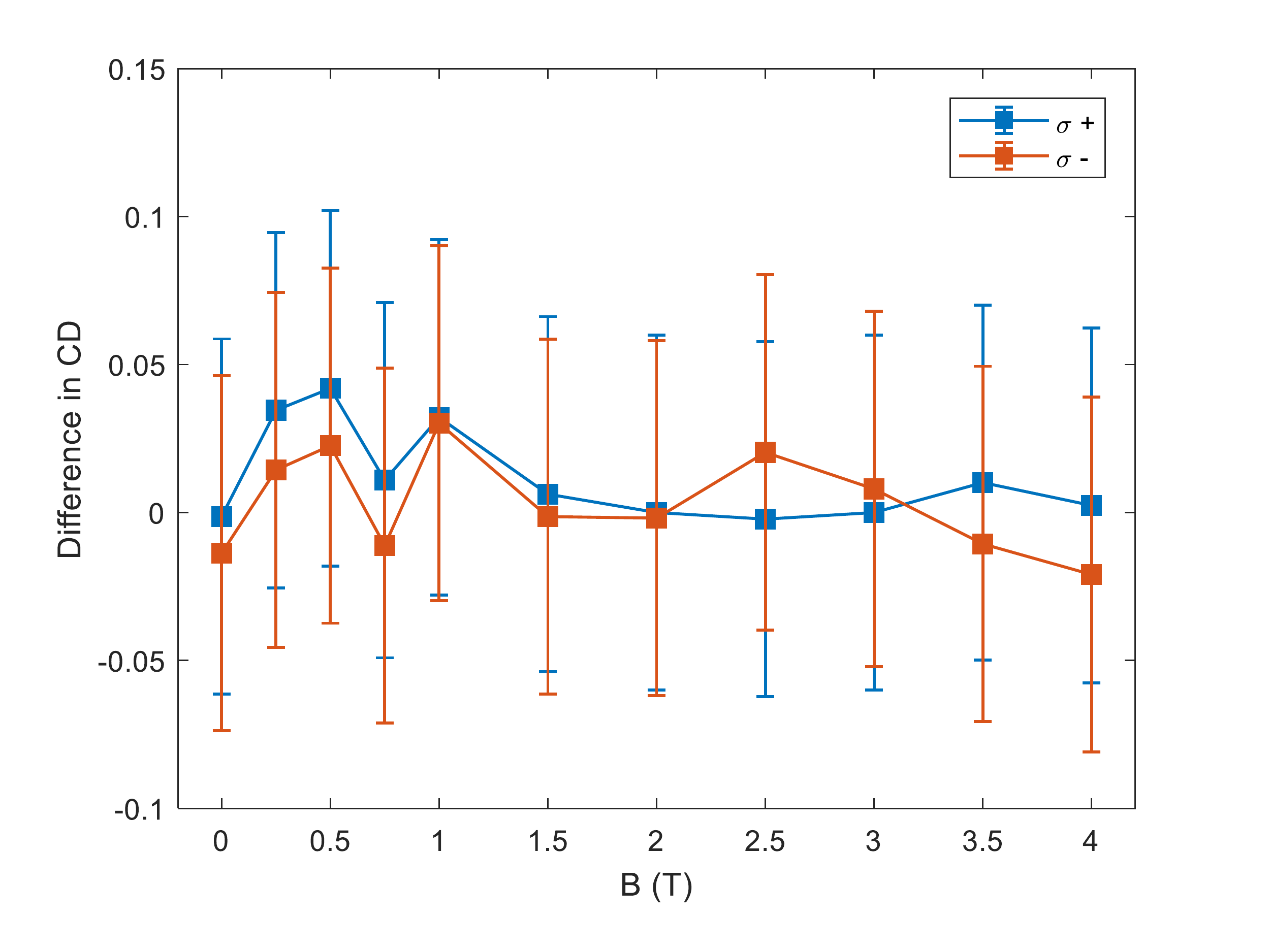} 
\end{center}
{\bf Figure S10: Difference of circular dichrosim between circular and linear excitation in B field of D2b red peak.} $\sigma^+$ and $\sigma^-$  excitations are shown in blue and red, respectively.
\end{figure}

\textbf {4. Polarization state and fidelity of phonon replica.}\\

In order to completely describe the polarization state of the emission, we estimate its density matrix elements by extracting the Stokes parameters $S_\mathrm{i}$, i = 0, 1, 2, 3. To this end, we measure the polarization of emission in both linear and circular basis as described in section~3. These measurements yield the intensities along horizontal (H), vertical (V), diagonal (D) and right circular (R) components from which we calculate $n_{â€¢0}$=$\dfrac{â€¢N}{â€¢2}$($\langle$H$\vert\widehat{â€¢\rho}\vert$H$\rangle$+$\langle$V$\vert\widehat{â€¢\rho}\vert$V$\rangle$), $n_{â€¢1}$=N$\langle\overline{â€¢D}\vert\widehat{â€¢\rho}\vert\overline{â€¢D}\rangle$, $n_{â€¢2}$=N$\langle$R$\vert\widehat{â€¢\rho}\vert$R$\rangle$ where the constant {\it N} is determined by light intensity and the efficiency of detector~\cite{Hecht20}. The Stokes parameters are defined as $S_{•0}$=2$n_{•0}$, $S_{•1}$=2($n_{•1}$-$n_{•0}$), $S_{•2}$=2($n_{•2}$-$n_{•0}$) and $S_{•3}$=2($n_{•3}$-$n_{•0}$) and ($\dfrac{•S_{•1}}{•S_{•0}}$, $\dfrac{•S_{•2}}{•S_{•0}}$, $\dfrac{•S_{•3}}{•S_{•0}}$) is called Stokes vector. For instance, Stokes vector, (0.005, -0.083, -0.023) of the phonon replica dot, D3b red peak is close to the unpolarized point in Poincar\'{e} sphere. The density matrix is defined as  $\widehat{•\rho}$=$\dfrac{•1}{•2}\sum_{i=0}^{3}\dfrac{•S_{•i}}{•S_{•0}}\widehat{•\sigma}_{•i}$, where $\widehat{•\sigma}_{•0}$ is identity operator and $\widehat{•\sigma}_{•1,2,3}$ are Pauli operators. In linear H and V basis, the unpolarized state’s density matrix is $\widehat{•\rho}_{•ideal}$=$\begin{bmatrix}\frac{1}{2} & 0\\0 &\frac{1}{2}\end{bmatrix}$ and from the experiment results, the density matrix of D3b red peak is $\widehat{•\rho}_{•exp}$=$\begin{bmatrix}0.500 & 0.014+0.007i\\0.014-0.007i &0.500 \end{bmatrix}$. Therefore, from the equation F=Tr($\sqrt{•\sqrt{•\widehat{•\rho}_{•ideal}}\widehat{•\rho}_{•exp}\sqrt{•\widehat{•\rho}_{•ideal}}}$), the fidelity of D3b red peak's emission is 0.999$\pm$0.016 . Fidelity close to unity indicates that the emitted light is unpolarized. For another phonon replica dot, D4b red peak, the density matrix is $\widehat{•\rho}_{•exp}$=$\begin{bmatrix}0.506 & 0.009-0.005i\\0.009+0.005i &0.494 \end{bmatrix}$ and fidelity is 0.999$\pm$0.013.\\

\begin{figure}[H]
\vskip 10pt
\begin{center}
\includegraphics[width=140mm]{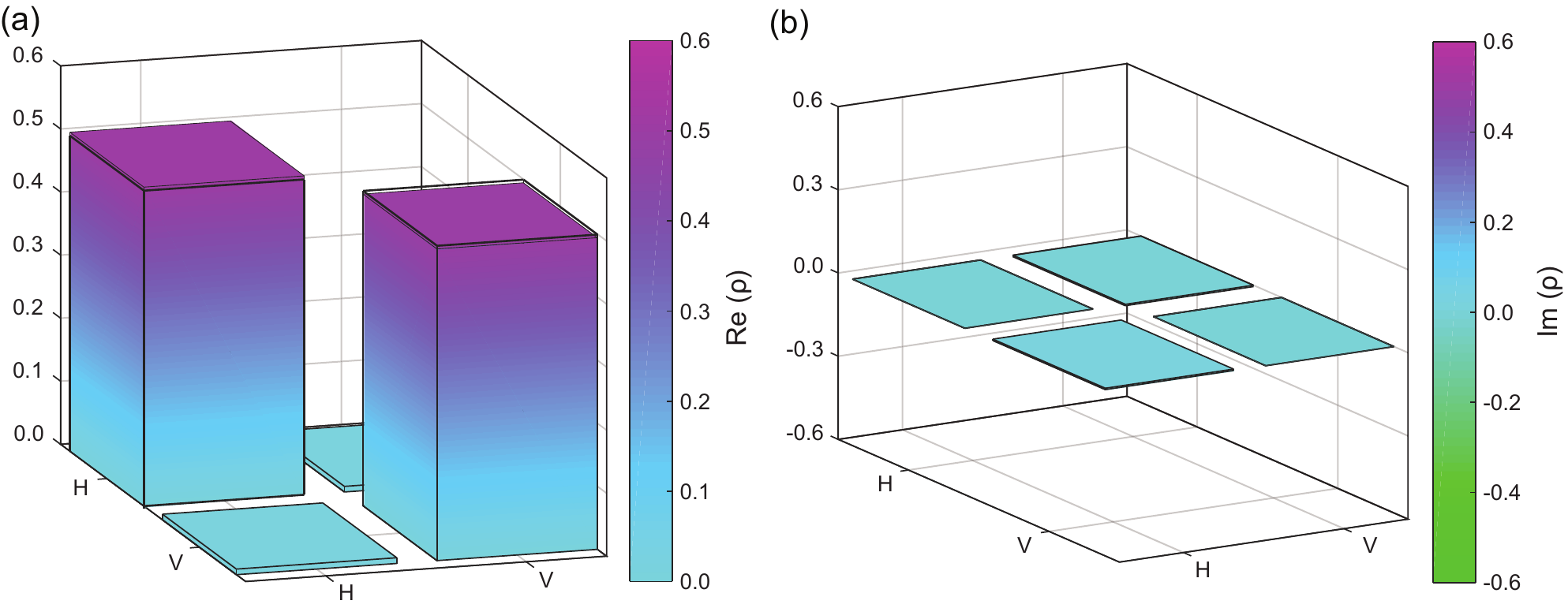}
\end{center}
{\bf Figure S11: Reconstructed density matrix of the polarization state of D4b red peak.} {\bf a,b} Real (a) and imaginary (b) parts of the density matrix elements. The hollow caps in (a) indicate the values (0.5) of matrix elements for a completely mixed state.
\end{figure}

\textbf {5. Symmetry analysis of $E''$ phonon mode in monolayer WSe$_2$}\\

Monolayer WSe$_2$ belongs to symmetry point group $D_{3h}^1$. The $E''$ phonon mode transforms like \textit{(x+iy)z} and \textit{(x-iy)z}, which corresponds to a pseudoangular momentum (PAM) of $+\hbar$ and $-\hbar$. As thickness increases, the counterpart of $E''$ mode in few-layer WSe$_2$ with even layer number (symmetry: $D_{3d}^3$) is $E_g$ mode. $E_g$ mode transforms like \textit{(x+iy)$^2$} and \textit{(x-iy)$^2$}, corresponding to PAM of $+2\hbar$ and $-2\hbar$. In few-layer WSe$_2$ with odd layer number (symmetry: $D_{3h}^1$), the $E''$ mode in monolayer splits into $E''$ and $E'$ modes. $E'$ mode transforms like \textit{(x+iy)$^2$} and \textit{(x-iy)$^2$} with PAM of $+2\hbar$ and $-2\hbar$. The phonon mode further changes to $E_{1g}$ in bulk 2H-phase WSe$_2$ (symmetry: $D_{6h}^4$), which transforms like \textit{(x+iy)z} and \textit{(x-iy)z} with PAM of $+\hbar$ and $-\hbar$ \cite{Loudon01}.

We can understand the selection rules by considering the threefold symmetry of the lattice. Due to conservation of momentum, the phonon with PAM of $+\hbar$($-\hbar$) can couple a $\sigma^{-}$($\sigma^{+}$) photon to a scattered $\sigma^{+}$($\sigma^{-}$)-polarized photon. This selection rule shows that the $E''$($E_g$/$E'$/$E_{1g}$) mode can be observed when incident and scattered light have opposite circular polarization, but disappears when the circular polarization is the same because conservation of momentum forbids the phonon mode to be observed~\cite{Drapcho17}. Helicity-resolved (non-resonance) Raman scattering measurements in Figure S12 confirms the analysis (2-5L), and our observation is in consistent with previous studies on other in-plane degenerate modes \cite{Drapcho17,Chen15,Yoshikawa17}. However, we also notice that the $E''$ mode disappears in monolayer \cite{Luo13,Kim17}. 

\begin{figure}[H]
\begin{center}
\includegraphics[width=100mm]{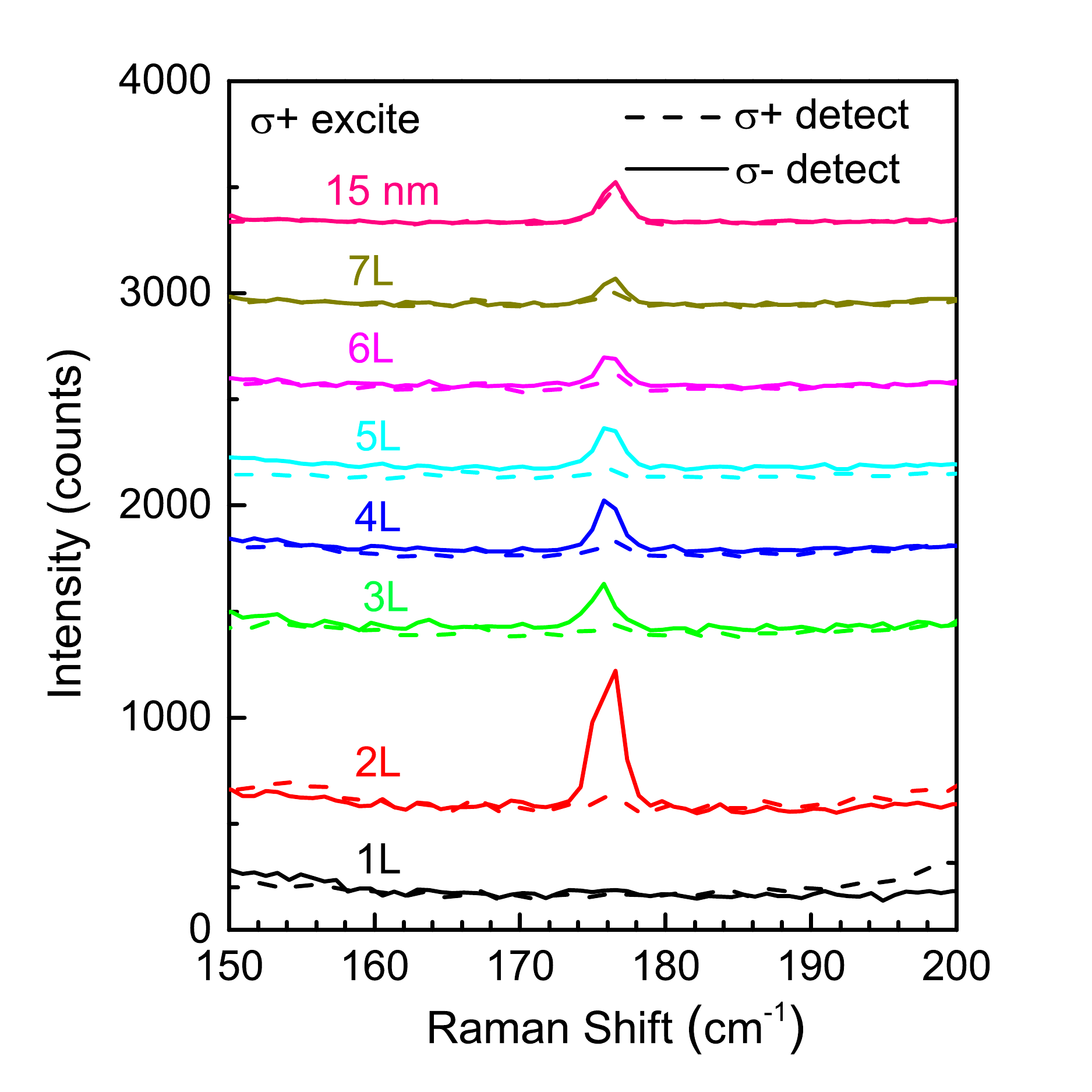}
\end{center}
{\bf Figure S12: Helicity-resolved Raman scattering of $E''$($E_g$/$E'$/$E_{1g}$) mode in WSe$_2$ showing reversal of helicity under non-resonant excitation.} The polarization of incident laser is $\sigma^+$. $E_g$/$E'$ mode appears in $\sigma^-$ polarized detection from 2-5L WSe$_2$, and has very small intensity when the detection configuration is in $\sigma^+$. While the selection rule is relaxed in thicker few-layers (6L and 7L) and bulk (15 nm). The $E''$ mode is not observed in 1L due to symmetry constraints. $\lambda_{\mathrm{exc}}$ = 441.6 nm and $P$ = 0.5 mW.
\end{figure}

The experimental observation of a given phonon mode in Raman spectroscopy depends on the symmetry selection rules as well as on the scattering geometry. A given phonon mode can be observed only when $\langle\widehat{e_i}\vert$\textbf{\textit{R}}$\vert\widehat{e_s}\rangle$ in nonzero. Here $\widehat{e_i}$ is the polarization vector of the incident light, and $\widehat{e_s}$ is that of the scattered light. The Raman tensor for the $E''$ phonon mode in monolayer WSe$_2$ is \textbf{\textit{R}} = $\begin{bmatrix}0 & 0 & a \\0 & 0 & b\\a& b & 0 \end{bmatrix}$~\cite{Luo13}. If we take linearly polarized light as basis, the $x$-polarized light and $y$-polarized light are $\begin{bmatrix}1 & 0 & 0 \end{bmatrix}$ and $\begin{bmatrix}0 & 1 & 0 \end{bmatrix}$, respectively. Circularly polarized light can be obtained through superposition: 1/$\sqrt{2}$$\begin{bmatrix}1 & i & 0 \end{bmatrix}$ and 1/$\sqrt{2}$$\begin{bmatrix}1 & - i & 0 \end{bmatrix}$. $\langle\widehat{e_i}\vert$\textbf{\textit{R}}$\vert\widehat{e_s}\rangle$ for $E''$ mode in monolayer WSe$_2$ is constantly zero on backscattering configuration regardless of the polarization of incident and scattered light. A QD in the sample lowers the symmetry from threefold to twofold {\it e.g}. from  $D_{3h}$ to $C_2$ group, as can be deduced by the anisotropy which results in a preferential axis for its polarization. The $E''$ phonon mode in monolayer can then have a nonzero Raman intensity.\\

As thickness increases, calculations on $\langle\widehat{e_i}\vert$\textbf{\textit{R}}$\vert\widehat{e_s}\rangle$ show that $E_g$/$E'$ mode should appear when incident and scattered light have opposite circular polarization, and disappear when the circular polarization is the same. The Raman tensor for $E_g$ mode in 2L (4L, 6L,...)WSe$_2$ is \textbf{\textit{R}} = $\begin{bmatrix}a & c & d \\c &-a & f\\d& f & 0 \end{bmatrix}$, and the Raman tensor for $E'$ mode in 3L (5L, 7L,...)WSe$_2$ is \textbf{\textit{R}} = $\begin{bmatrix}a & c & 0\\c &-a & 0\\0& 0& 0 \end{bmatrix}$~\cite{Luo13}.  Raman scattering measurements on 2-5L confirms the above analysis with polarization, ($I_+$-$I_-$)/($I_+$+$I_-$), of almost -100$\%$ (Figure S12). We note that the selection rule is relaxed in thicker few-layers (6L and 7L) and bulk (15 nm). The Raman tensor for $E_{1g}$ mode in bulk is same as $E''$ in monolayer, which indicates that the $E_{1g}$ mode is not supposed to appear. However, $E_{1g}$ mode shows up in both $\sigma^+$ and $\sigma^-$ detection configurations with similar intensities.\\

\clearpage

\textbf{6. Photoluminescence excitation spectroscopy (PLE) of D2a, D3a and D4a doublets.}

We conducted the PLE measurements with laser tuned from the low energy to high energy side. For D3a and D4a doublets, a prominent peak with energy $\sim$53 meV above the emission shows up in the PLE spectra, which could be the 2$s$ state of the QD~\cite{Tonndorf15, Wang15}.

\begin{figure}[H]
\begin{center}
\includegraphics[width=160mm]{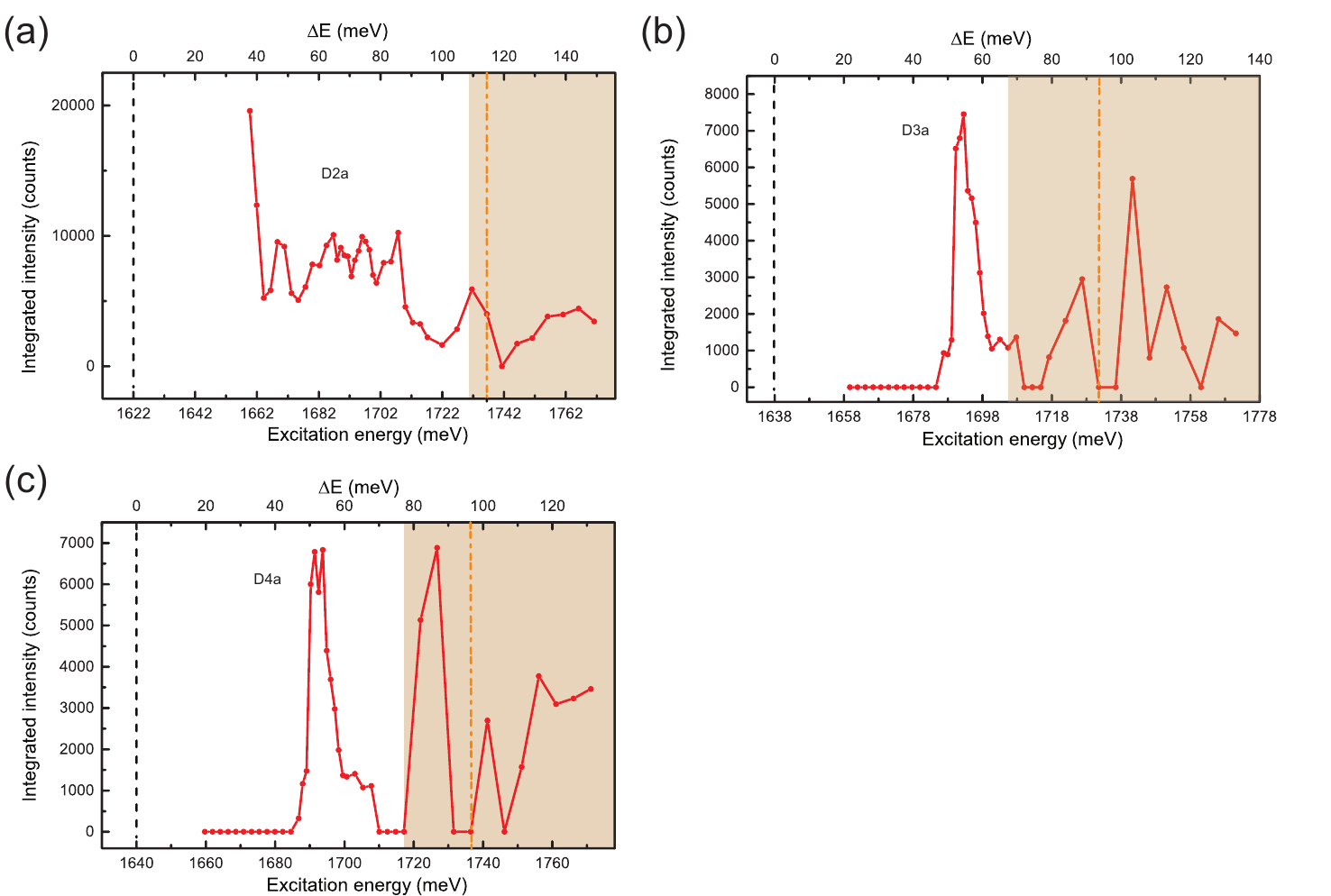}\\
\end{center}
{\bf Figure S13: PLE of D2a, D3a and D4a doublets.} Black dash line indicates the emission energy of the doublets, and orange dash-dot line shows the energy of the free exciton $X^0$. The PLE measurements were carried out with laser tuned from the low energy to high energy side. As excitation energy increases, defect states are populated with broad background shown in the spectra. Thus, the extracted intensities in these regions (brown shaded regions) cannot fully reflect the PLE profile. For D3a and D4a doublets, a prominent peak with energy $\sim$53 meV above the emission shows up in the PLE spectra, which could be the 2$s$ state of the QD~\cite{Tonndorf15, Wang15}.
\end{figure}

{}

\end{document}